\documentclass[pra,aps,showpacs,onecolumn,twoside,superscriptaddress]{revtex4}

\usepackage{amsmath,amsfonts,amssymb,caption,color,epsfig,graphics,graphicx,hyperref,latexsym,mathrsfs,revsymb,theorem,url,verbatim,epstopdf}

\newtheorem{definition}{Definition}
\newtheorem{proposition}[definition]{Proposition}
\newtheorem{lemma}[definition]{Lemma}

\newtheorem{theorem}[definition]{Theorem}
\newtheorem{corollary}[definition]{Corollary}
\newtheorem{conjecture}[definition]{Conjecture}

\newtheorem{remark}[definition]{Remark}
\newtheorem{example}[definition]{Example}
\newtheorem{question}[definition]{Question}

\def\squareforqed{\hbox{\rlap{$\sqcap$}$\sqcup$}}
\def\qed{\ifmmode\squareforqed\else{\unskip\nobreak\hfil
\penalty50\hskip1em\null\nobreak\hfil\squareforqed
\parfillskip=0pt\finalhyphendemerits=0\endgraf}\fi}
\def\endenv{\ifmmode\;\else{\unskip\nobreak\hfil
\penalty50\hskip1em\null\nobreak\hfil\;
\parfillskip=0pt\finalhyphendemerits=0\endgraf}\fi}
\newenvironment{proof}{\noindent \textbf{{Proof.~} }}{\qed}
\def\Dbar{\leavevmode\lower.6ex\hbox to 0pt
{\hskip-.23ex\accent"16\hss}D}
\makeatletter
\def\url@leostyle{%
  \@ifundefined{selectfont}{\def\UrlFont{\sf}}{\def\UrlFont{\small\ttfamily}}}
\makeatother
\def\bcj{\begin{conjecture}}
\def\ecj{\end{conjecture}}
\def\bcr{\begin{corollary}}
\def\ecr{\end{corollary}}
\def\bd{\begin{definition}}
\def\ed{\end{definition}}
\def\bea{\begin{eqnarray}}
\def\eea{\end{eqnarray}}
\def\bem{\begin{enumerate}}
\def\eem{\end{enumerate}}
\def\bex{\begin{example}}
\def\eex{\end{example}}
\def\bim{\begin{itemize}}
\def\eim{\end{itemize}}
\def\bl{\begin{lemma}}
\def\el{\end{lemma}}
\def\bpf{\begin{proof}}
\def\epf{\end{proof}}
\def\bpp{\begin{proposition}}
\def\epp{\end{proposition}}
\def\bqu{\begin{question}}
\def\equ{\end{question}}
\def\br{\begin{remark}}
\def\er{\end{remark}}
\def\bt{\begin{theorem}}
\def\et{\end{theorem}}

\def\btb{\begin{tabular}}
\def\etb{\end{tabular}}

\newcommand{\nc}{\newcommand}

\def\a{\alpha}
\def\b{\beta}
\def\g{\gamma}
\def\d{\delta}
\def\e{\epsilon}

\def\r{\rho}
\def\s{\sigma}

\def\ph{\varphi}

\def\ps{\psi}
\def\o{\omega}

\def\G{\Gamma}

\def\L{\Lambda}

\def\P{\Pi}

\def\Ps{\Psi}

 \nc{\bbA}{\mathbb{A}} \nc{\bbB}{\mathbb{B}} \nc{\bbC}{\mathbb{C}}
 \nc{\bbD}{\mathbb{D}} \nc{\bbE}{\mathbb{E}} \nc{\bbF}{\mathbb{F}}
 \nc{\bbG}{\mathbb{G}} \nc{\bbH}{\mathbb{H}} \nc{\bbI}{\mathbb{I}}
 \nc{\bbJ}{\mathbb{J}} \nc{\bbK}{\mathbb{K}} \nc{\bbL}{\mathbb{L}}
 \nc{\bbM}{\mathbb{M}} \nc{\bbN}{\mathbb{N}} \nc{\bbO}{\mathbb{O}}
 \nc{\bbP}{\mathbb{P}} \nc{\bbQ}{\mathbb{Q}} \nc{\bbR}{\mathbb{R}}
 \nc{\bbS}{\mathbb{S}} \nc{\bbT}{\mathbb{T}} \nc{\bbU}{\mathbb{U}}
 \nc{\bbV}{\mathbb{V}} \nc{\bbW}{\mathbb{W}} \nc{\bbX}{\mathbb{X}}
 \nc{\bbZ}{\mathbb{Z}}

 \nc{\bA}{{\bf A}} \nc{\bB}{{\bf B}} \nc{\bC}{{\bf C}}
 \nc{\bD}{{\bf D}} \nc{\bE}{{\bf E}} \nc{\bF}{{\bf F}}
 \nc{\bG}{{\bf G}} \nc{\bH}{{\bf H}} \nc{\bI}{{\bf I}}
 \nc{\bJ}{{\bf J}} \nc{\bK}{{\bf K}} \nc{\bL}{{\bf L}}
 \nc{\bM}{{\bf M}} \nc{\bN}{{\bf N}} \nc{\bO}{{\bf O}}
 \nc{\bP}{{\bf P}} \nc{\bQ}{{\bf Q}} \nc{\bR}{{\bf R}}
 \nc{\bS}{{\bf S}} \nc{\bT}{{\bf T}} \nc{\bU}{{\bf U}}
 \nc{\bV}{{\bf V}} \nc{\bW}{{\bf W}} \nc{\bX}{{\bf X}}
 \nc{\bZ}{{\bf Z}}

\nc{\cA}{{\cal A}} \nc{\cB}{{\cal B}} \nc{\cC}{{\cal C}}
\nc{\cD}{{\cal D}} \nc{\cE}{{\cal E}} \nc{\cF}{{\cal F}}
\nc{\cG}{{\cal G}} \nc{\cH}{{\cal H}} \nc{\cI}{{\cal I}}
\nc{\cJ}{{\cal J}} \nc{\cK}{{\cal K}} \nc{\cL}{{\cal L}}
\nc{\cM}{{\cal M}} \nc{\cN}{{\cal N}} \nc{\cO}{{\cal O}}
\nc{\cP}{{\cal P}} \nc{\cQ}{{\cal Q}} \nc{\cR}{{\cal R}}
\nc{\cS}{{\cal S}} \nc{\cT}{{\cal T}} \nc{\cU}{{\cal U}}
\nc{\cV}{{\cal V}} \nc{\cW}{{\cal W}} \nc{\cX}{{\cal X}}
\nc{\cZ}{{\cal Z}}

\nc{\hA}{{\hat{A}}} \nc{\hB}{{\hat{B}}} \nc{\hC}{{\hat{C}}}
\nc{\hD}{{\hat{D}}} \nc{\hE}{{\hat{E}}} \nc{\hF}{{\hat{F}}}
\nc{\hG}{{\hat{G}}} \nc{\hH}{{\hat{H}}} \nc{\hI}{{\hat{I}}}
\nc{\hJ}{{\hat{J}}} \nc{\hK}{{\hat{K}}} \nc{\hL}{{\hat{L}}}
\nc{\hM}{{\hat{M}}} \nc{\hN}{{\hat{N}}} \nc{\hO}{{\hat{O}}}
\nc{\hP}{{\hat{P}}} \nc{\hR}{{\hat{R}}} \nc{\hS}{{\hat{S}}}
\nc{\hT}{{\hat{T}}} \nc{\hU}{{\hat{U}}} \nc{\hV}{{\hat{V}}}
\nc{\hW}{{\hat{W}}} \nc{\hX}{{\hat{X}}} \nc{\hZ}{{\hat{Z}}}

\nc{\hn}{{\hat{n}}}

\def\dim{\mathop{\rm Dim}}

\def\lin{\mathop{\rm span}}

\def\max{\mathop{\rm max}}
\def\min{\mathop{\rm min}}

\def\rank{\mathop{\rm rank}}

\def\tr{\mathop{\rm Tr}}

\def\sn{\mathop{\rm SN}}

\def\snmax{\mathop{\rm SN_{\max}}}
\def\snmin{\mathop{\rm SN_{\min}}}

\def\jsn{\mathop{\rm JSN}}

\def\dg{\dagger}

\def\op{\oplus}
\def\ox{\otimes}

\def\ra{\rightarrow}

\def\su{\subset}
\def\sue{\subseteq}

\newcommand{\bra}[1]{\langle#1|}
\newcommand{\ket}[1]{|#1\rangle}
\newcommand{\proj}[1]{| #1\rangle\!\langle #1 |}
\newcommand{\ketbra}[2]{|#1\rangle\!\langle#2|}
\newcommand{\braket}[2]{\langle#1|#2\rangle}

\def\Dbar{\leavevmode\lower.6ex\hbox to 0pt
{\hskip-.23ex\accent"16\hss}D}

\begin{document}
\title{Schmidt number of bipartite and multipartite states under local projections}

\author{Lin Chen}
\email{linchen@buaa.edu.cn} 
\affiliation{School of Mathematics and Systems Science, Beihang University, Beijing 100191, China}
\affiliation{International Research Institute for Multidisciplinary Science, Beihang University, Beijing 100191, China}

\author{Yu Yang}
\email{a0086285@nus.edu.sg}
\affiliation{Department of Mathematics, National University of Singapore, 10 Lower Kent Ridge Road, Singapore 119076, Republic of Singapore}

\author{Wai-Shing Tang}
\email{mattws@nus.edu.sg}
\affiliation{Department of Mathematics, National University of Singapore, 10 Lower Kent Ridge Road, Singapore 119076, Republic of Singapore}

\date{\today}


\begin{abstract}
The Schmidt number is a fundamental parameter characterizing the properties of quantum states, and the local projections are a fundamental operation in quantum physics. We investigate the relation between the Schmidt numbers of bipartite states and their projected states. We show that there exist bipartite positive-partial-transpose (PPT) entangled states of any given Schmidt number. We further construct the notion of joint Schmidt number for multipartite states, and its relation with the Schmidt number of bipartite reduced density operators.

\end{abstract}

\maketitle

\tableofcontents

\section{Introduction}
\label{sec:int}

The Schmidt number is a parameter characterizing quantum states. A quantum state is entangled if and only if its Schmidt number is greater than one.
Entangled states play the fundamental role in quantum-information applications such as quantum computing and cryptography. One quantum state $\r$ is converted into another state $\s$ under the physical environment of local operations and classical communications (LOCC). In spite of the complex mathematical configuration of LOCC, the most basic operation in LOCC is the local projection $P$. Mathematically we have $\r\ra\s=(I_A\ox P_B)\r(I_A\ox P_B)$. In this process the Schmidt number is non-increasing, and the decrease of Schmidt number is decided by the local projection. 
In this paper we begin by recalling the Schmidt number in Definition \ref{df:sn}, and the notion of birank. Then we construct the notion of bi-Schmidt number in Eq.  \eqref{eq:bsn}. We further provide the upper bound of entanglement of formation of quantum states in terms of the Schmidt number in \eqref{eq:efr}. The bound is saturated when the states are antisymmetric two-qubit states. Next we recall the definition of direct sum and tensor product of two quantum states, and obtain a few preliminary results in Lemma \ref{le:dsum} and \ref{le:channel}.
The entanglement of the tensor product of two quantum states is invesitgated in Lemma \ref{le:tensorof2}. Next we recall the the positive and copositive maps in Definition \ref{df:kpositive} and \ref{df:cp}. As an application, we show in Lemma \ref{le:SNstb} that for any bipartite states $\r$ and $\s$ with $\sn(\s)\leq\sn(\r)$, the Schmidt number of the perturbation $\r+\epsilon\s$ remains $\sn(\r)$ for sufficiently small $\epsilon>0$.

The main result of this paper is as follows. We will investigate how the projection influences the Schmidt number of both bipartite and multipartite states. For bipartite states the investigation is carried out in Lemma \ref{le:proj} and \ref{le:sr2copy}. As an application we show that every positive-partial-transpose (PPT) entangled $\r$ is of Schmidt number 2 in Corollary \ref{cr:3x3}. It provides an alternative proof for a conjecture in \cite{sbl2001}. We further show that the projected state $\s$ can reach any integer smaller than the Schmidt number of $\r$ in Lemma \ref{le:sym}. As an application of this result, we show that there exist bipartite PPT entangled states of any given integer in Theorem \ref{thm:sr}. This is based on the preliminary results developed in Lemma \ref{le:>2} and Proposition \ref{pp:snrho>n}. We also investigate when an entangled state can be projected onto a separable state in terms of their rank. For multipartite states, we introduce the notion of expansion and coarse graining respectively in Definition \ref{df:expansion} and \ref{df:combine}. We investigate their relation to the Schmidt number of bipartite reduced density operators in Theorem \ref{le:expansion} and Lemma \ref{le:combine}. We further construct the notion of joint Schmidt number for multipartite states in Definition \ref{df:biSR} and \ref{df:JSN}. We also restrict the joint Schmidt number of a multipartite pure state by the Schmidt numbers of its bipartite reduced density operators in Theorem \ref{le:jsn}. As an application, we show in Lemma \ref{le:rank4} that any multipartite entangled PPT state with Schmidt number at least 3 when regarded as bipartite states, has rank at least 5. 


The rest of the paper is organized as follows. In Sec. \ref{sec:pre} we introduce the preliminary definitions, notations and facts used in the paper. They include the Schmidt number in Sec. \ref{subsec:sn}, the positive map in Sec. \ref{subsec:pm} and linear algebra in Sec. \ref{subsec:la}. In Sec. \ref{sec:main} we show that there exist bipartite PPT entangled states of any given Schmidt number. Next we introduce the notion of expansion and coarse graining of multipartite states in terms of the Schmidt number respectively in Sec. \ref{subsec:exp} and \ref{subsec:coa}. We further present the joint Schmidt number for multipartite states, and their relation to bipartite reduced density operator in Sec. \ref{subsec:jsn}. 

\section{Preliminaries}
\label{sec:pre}

Let $\cH=\cH_A\ox\cH_B$ be the bipartite Hilbert space  with $\dim\cH_A=M$ and  $\dim\cH_B=N$. Since the case $M=1$ or $N=1$ is trivial, we assume $2\le M\le N$. We say that $\r$ is a $M\times N$ state when $\rank\r_A=M$ and $\rank\r_B=N$. We shall work with bipartite quantum states $\r$ on $\cH$. We shall write $I_k$ for the identity $k\times k$ matrix.We denote by $\cR(\r)$ and $\ker \r$ the range and kernel of a linear map
$\r$, respectively. From now on, unless stated otherwise, the states will not
be normalized. We shall denote by $\{\ket{i}_A:i=0,\ldots,M-1\}$ and
$\{\ket{j}_B:j=0,\ldots,N-1\}$ o. n. bases of $\cH_A$ and $\cH_B$,
respectively. The partial transpose of $\r$ w. r. t. the system $A$ is defined as $\rho^{\G}:=\sum_{i,j}\ketbra{j}{i}\ox\bra{i}\r\ket{j}$. We say that $\r$ is PPT if $\r^{\G}\ge0$. Otherwise $\r$ is NPT, i.e., $\r^{\G}$ has at least one negative eigenvalue. We say that two bipartite states $\r$ and $\s$ are equivalent
under SLOCC if there exists an invertible local operator
(ILO) $A\ox B$ such that $\r=(A^\dg \ox B^\dg) \s (A \ox B)$ \cite{dvc2000}. In particular, they are locally equivalent when $A$ and $B$ are unitary matrices.
It is easy to see
that any ILO transforms distillable, PPT, entangled, or separable state into the
same kind of states. We shall often use ILOs to simplify the density
matrices of states. A subspace which contains no product state, 
is referred to as a completely entangled subspace (CES).

In the following subsections, we respectively introduce the Schmidt number, the positive map, and a few results from linear algebra. In Sec. \ref{subsec:sn}, we review the Schmidt number in Definition \ref{df:sn}, construct the notion of bi-Schmidt number in \eqref{eq:bsn} and introduce the direct sum and tensor product of two bipartite states.  In Sec. \ref{subsec:pm}, we review the positive and copositive map in Definition \ref{df:kpositive}, and the completely positive and copositive map in Definition \ref{df:cp}. We further review the reduction map and investigate a family of $k$-positive map. In Sec. \ref{subsec:la} we review and construct a few results on linear algebra. Lemma \ref{le:th00} shows a corollary in terms of maximally entangled states, when a bipartite state has a given Schmidt number.

\subsection{Schmidt number}
\label{subsec:sn}

In this subsection we review the definition of Schmidt number \cite{th00} and its physical meanings. Then we construct the notion of bi-Schmidt number for PPT states. We also review the B-direct sum of quantum states, entanglement of formation, and quantum channel, and their relation to the Schmidt number.
\bd
\label{df:sn}
A bipartite density matrix $\r$􏰓 has Schmidt number $\sn(\r)=k$ if (i) for any decomposition $\{p_i\ge0, \ket{\ps_i}\}$ of $\r$􏰓,  at least one of the vectors $\ket{\ps_i}$􏰃 has Schmidt rank at least $k$ and (􏰕ii)􏰀 there exists a decomposition of􏰓 $\r$ with all vectors $􏰂􏰥\ket{\ps_i}$􏰄􏰃 of Schmidt rank at most $k$.
\ed
For example the $M\times N$ pure state has Schmidt number $M$.
Another example is that the two-qubit mixed state $\r=\proj{\a}+\proj{00}$ where $\ket{\a}=\ket{00}+\ket{11}$ has Schmidt number two. To understand this fact, we assume that $\r=\sum_i p_i \proj{\ps_i}$ as an arbitrary decomposition of $\r$. 
As we shall see in Lemma \ref{le:sd}, one can obtain that there is always some $\ket{\ps_i}$ of tensor rank two. Then Definition \ref{df:sn} shows that $\sn(\r)=2$.
It further implies that the Schmidt number of bipartite states does not increase under LOCC. So the Schmidt number is an entanglement monotone for bipartite states.
For simplicity we denote $\sn(\r)$ as the Schmidt number of $\r$. 
Suppose $\r=p\a+(1-p)\b$ is a quantum state, $\a,\b$ are two states, and $p\in(0,1)$. It is known that for some $\r$ we have $\sn (\r) < p \sn (\a) + (1-p)\sn(\b)$, e.g., $\a=2(\ket{00}+\ket{11})(\bra{00}+\bra{11})+(\ket{00}-\ket{11}+\ket{22})(\bra{00}-\bra{11}+\bra{22})$, $\b=(\ket{00}-\ket{11}-\ket{22})(\bra{00}-\bra{11}-\bra{22})$ and $p=1/2$. 
On the other hand, suppose that $\r=(1-p)\s+pI$ is a quantum state and $\s$ has Schmidt number two. By Lemma \ref{le:SNstb} if $p$ is small enough then $\r$ has $\sn(\r)=2$.
Then we have $\sn (\r) > (1-p) \sn (\s) + p \sn(I)$. The above two examples imply that the Schmidt number is neither convex nor concave, although Schmidt number is an entanglement monotone. This is different from many known entanglement monotones  in quantum information, e.g., the entanglement of formation is convex \cite{bds96}.
Meanwhile, the Schmidt number of the state $\r=p\proj{\a}+(1-p)\proj{\b}$ may rely on $p$.  An example is $\ket{\a}={1\over\sqrt2}(\ket{00}+\ket{11})$ and $\ket{\b}={1\over\sqrt2}(\ket{00}-\ket{11})$. We can easily show that $\r$ is separable if and only if $p=1/2$, and entangled otherwise. This physical phenomenon mathematically corresponds to the change of Schmidt number of $\r$ between $1$ and $2$. 

To apply the Schmidt number to PPT states $\r$, we recall the notion of birank. It has been used to investigate the two-qutrit PPT entangled states of rank four \cite{cd11}. 
We denote the pair of integers $(\rank\r,\rank\r^\G)$ as the \textit{birank} of $\r$. Similar to the birank, we denote the pair of integers 
\bea
\label{eq:bsn}
(\sn(\r),\sn(\r^\G))
\eea 
as the \textit{bi-Schmidt number}, namely the BSN of $\r$. Unlike the birank, the BSN is defined for PPT states only because the Schmidt number is defined only for quantum states. Below is an application of BSN. The proof is by the fact that the partial transpose of a separable state is still separable.
\bl
\label{le:sn12}
If $\r$ is a PPT state and $\sn(\r),\sn(\r^\G)\in\{1,2\}$ then $\sn(\r)=\sn(\r^\G)$.
\el

Here is another application of the Schmidt number. We refer to $E_f(\r)$ as the entanglement of formation (EOF) of the state $\r$. It is a fundamental entanglement measure for quantum states and has been widely investigated in the past years \cite{bds96}. However the estimation of the bound of EOF has been an involved problem. Definition \ref{df:sn} implies that 
\bea
\label{eq:efr}
E_f(\r)\le \log_2 \sn(\r),
\eea
i.e., an upper bound of EOF of $\r$ is $\log_2 \sn(\r)$ ebits. It is known that any quantum state in the 3-dimensional antisymmetric subspace $\cA$ is locally equivalent to a two-qubit maximally entangled state. So $E_f(\r)=1$ ebit and $\sn(\r)=2$ when $\cR(\r)\sue\cA$. It implies that the equality in \eqref{eq:efr} holds when $\cR(\r)\sue\cA$. In this sense, the EOF of antisymmetric states is analytically characterized by their Schmidt number.

Next we investigate the Schmidt number of the collective use of two quantum states. For this purpose we introduce two notions from quantum information. The first notion is the direct sum of two spaces. It plays an important role in many quantum-information problems such as the distillability problem \cite{cd16dis} and bipartite unitary operations \cite{cy14,cy14ap,cy15}.
We shall denote $V\op W$ as the ordinary direct sum of two matrices $V$ and $W$, and $V\op_B W$ as the direct sum of $V$ and $W$ from the $B$ side (called ``$B$-direct sum''). In the latter case, $V$ and $W$ respectively act on two subspaces $\cH_A\ox\cH'_B$ and $\cH_A\ox\cH''_B$ such that $\cH_B'\perp\cH_B''$. We shall denote the tensor product of two bipartite states $\r_{A_1B_1}$ and $\s_{A_2B_2}$ as another bipartite state of the system $A_1A_2$ and $B_1B_2$.

The second notion from quantum information is the combination of different systems. Let $\r_{A_iB_i}$ be an $M_i\times N_i$ state of rank $r_i$ acting
on the Hilbert space $\cH_i=\cH_{A_i}\ox\cH_{B_i}$, $i=1,2$. Suppose
$\r$ of systems $A_1,A_2$ and $B_1,B_2$ is a state acting on the
Hilbert space
$\cH_1\ox\cH_2=\cH_{A_1}\ox\cH_{B_1}\ox\cH_{A_2}\ox\cH_{B_2}$. By
switching the two middle factors, we can consider $\r$ as a
\textit{composite} bipartite state acting on the Hilbert space
$\cH_A\ox\cH_B$ where $\cH_A=\cH_{A_1}\ox\cH_{A_2}$ and
$\cH_B=\cH_{B_1}\ox\cH_{B_2}$. In that case we shall write
$\r=\r_{A_1A_2:B_1B_2}$. Let $\tr_{A_1B_1}\r=\r_{A_2B_2}$ and
$\tr_{A_2B_2}\r=\r_{A_1B_1}$. So $\r$ is an $M_1M_2\times N_1N_2$
state of rank not larger than $r_1r_2$. In particular for the
\textit{tensor product} $\r=\r_{A_1B_1}\ox\r_{A_2B_2}$, it is easy
to see that $\r$ is an $M_1M_2\times N_1N_2$ state of rank $r_1r_2$.
The above definition can be easily generalized to the tensor product 
of $N$ states $\r_{A_iB_i},i=1,\ldots,N$. They form a bipartite
state on the Hilbert space
$\cH_{A_1,\cdots,A_N}\ox\cH_{B_1,\cdots,B_N}$.
For simplicity we denote the system $A$ as $A_1,\cdots,A_N$ and denote $B$ as $B_1,\cdots,B_N$. 
For example, it is known that $\sn(\r^{\ox2}) \in [ \sn(\r),\sn(\r)^2]$, and $\sn(\r^{\ox2})$ may reach any integer in the interval $ [ \sn(\r),\sn(\r)^2]$ when $\sn(\r)=M$. An example is the two-qubit isotropic state \cite[Fig. 1]{th00}. 
Now we have
\bl
\label{le:dsum}
Suppose $\r=\a\op_B\b$ where $\a$ and $\b$ are both bipartite quantum states. Then
\\
(i) $\sn(\r)=\max\{\sn(\a),\sn(\b)\}.$
\\
(ii) $\sn(\r^{\ox n})=\max\{\sn(\a\ox\cdots\ox\a),
\sn(\a\ox\cdots\ox\a\ox\b),
\cdots,
\sn(\b\ox\a\ox\cdots\ox\a),
\cdots,
\sn(\b\ox\cdots\ox\b)\}$.
\el
\bpf
(i) By definition we have $\sn(\r)\le\max\{\sn(\a),\sn(\b)\}.$ On the other hand we can project $\r$ onto $\a$ and $\b$ by local projectors. 
Since the Schmidt number is an entanglement monotone we have $\sn(\r)\ge\max\{\sn(\a),\sn(\b)\}.$

(ii) The assertion follows from (i).
This completes the proof.
\epf

We generalize the Lemma as follows. It is known that any quantum physical operation can be expressed as a completely positive (CP) map $\L(\r) := \sum_i P_i \r P_i^\dg$ where $\sum_i P_i^\dg P_i \le I$. If the equality holds then the operation is a completely positive trace-preserving (CPTP) map, namely a quantum channel. We  construct the relation between quantum operation and Schmidt number.
\bl
\label{le:channel}
Suppose $\r$ is a bipartite state, and $\L(\cdot)=\sum_i P_i (\cdot) P_i^\dg$ is a quantum operation such that $(I_A\ox\L)\r=\r$. Then $\sn(\r)=\max_i\{\sn(\r_i)\}$ where $\r_i=(I_A\ox P_i) \r (I_A\ox P_i^\dg)$.
\el
\bpf
By definition we have $\sn(\r)\le\max_i\{\sn(\r_i)\}$. 
Since the Schmidt number is an entanglement monotone we have $\sn(\r)\ge\max_i\{\sn(\r_i)\}$.
This completes the proof.
\epf

If the channel is $\L(\cdot)=P(\cdot)P^\dg + (I-P)(\cdot)(I-P^\dg)$ where $P$ is a projector, then Lemma \ref{le:channel} reduces to Lemma \ref{le:dsum}. Finding out the states $\r$ satisfying the hypothesis of Lemma \ref{le:channel} is an interesting question. For example, we can assume $\r$ as the quantum-classical separable state $\r=\sum_i p_i \r_i \ox \proj{i}$ \cite{ccm11}.

The following Lemma investigates the entanglement of the tensor product of two quantum states. 
\bl
\label{le:tensorof2}
Let the integers $m_1,n_1,m_2,n_2\in\{2,3\}$, $m_1+n_1<6$ and $m_2+n_2<6$.
Suppose $\rho_1$ and $\rho_2$ are $m_1\times n_1$ and $m_2\times n_2$ states in $\cH_{A_1}\ox\cH_{B_1}$ and $\cH_{A_2}\ox\cH_{B_2}$, respectively. $\rho_1\ox\rho_2$ is a bipartite state w.r.t the bi-partition $A_1A_2:B_1B_2$.\\
(i) If either of the two states $\rho_1$ and $\rho_2$ is entangled, then $\rho_1\ox\rho_2$ is a NPT state.\\
(ii) Conversely, if $\rho_1\ox\rho_2$ is a PPT state, then both $\rho_1$ and $\rho_2$ are separable states.
\el
\bpf
(i) Assume that $\rho_1$ is entangled. It follows from the Peres-Horodecki criterion \cite{peres1996}, that the least eigenvalues of $\rho_1^{\G_{A_1}}$ is negative and the largest eigenvalues of $\rho_1^{\G_{A_2}}$ is positive. Since the eigenvalues of $(\rho_1\ox\rho_2)^{\G_{A_1A_2}}=\rho_1^{\G_{A_1}}\ox\rho_2^{\G_{A_2}}$ are the pairwise products of eigenvalues of $\rho_1^{\G_{A_1}}$ and $\rho_2^{\G_{A_2}}$, there exists a negative eigenvalue in the spectrum of $(\rho_1\ox\rho_2)^{\G_{A_1B_1}}$.\\
(ii) follows (i) immediately. This completes the proof.
\epf

The Lemma shows that the entanglement of the tensor product implies the entanglement of at least one state in the tensor product. On the other hand,
if $\r_1+\r_2$ is a separable state then $\r_1$ and $\r_2$ may be both entangled. An example is $\r_1=\proj{\a_{+}}$ and $\r_2=\proj{\a_{-}}$ where $\ket{\a_{\pm}}=\ket{11}\pm\ket{22}$. This is different from (ii) which works for the tensor product of two states. Moreover if we want to construct PPT entangled states using the tensor product of two PPT entangled states by Lemma \ref{le:tensorof2}, then $\rho_1$ and $\rho_2$ have to be $M\times N$ PPT entangled states where $M,N\geq3$. 


As another application of Schmidt rank, we introduce a subspace containing only highly entangled states  \cite{cmw08}.
\bd
\label{d:k-CES}
A subspace of $\mathbb{C}^m\ox \mathbb{C}^n$ is said to be a $k$-CES $(k\leq min\{m,n\})$ if it contains no nonzero Schmidt rank l vectors for $l\leq k$.
\ed

For example, if the range of a bipartite state is 1-completely entangled then the state is entangled. This is how the PPT entangled states by unextendible product bases are constructed \cite{BDM+99}. The definition of Schmidt number implies

\bl
If $\r$ is a bipartite quantum state whose $R(\r)$ is a $k$-CES, then $\sn(\r)\geq k+1$.
\el

The Lemma gives a sufficient condition such that $\r$ is entangled. The condition is not necessary. An example is the two-qubit state $\proj{00}+(\ket{00}+\ket{11})(\bra{00}+\bra{11})$. One can easily show that the state is entangled and its range is not 1-completely entangled.
Hitherto most results shows that estimating the Schmidt number is a hard problem. The following result from \cite{th00} provides a method for the estimation in terms of the maximally entangled states.
\bl
\label{le:th00}
For any density matrix $\r$ with $M=N$ and Schmidt number $k$, we have
\bea
\max_{\Ps_M} \bra{\Ps_M} \r \ket{\Ps_M} \le {k \over N},
\eea
where we maximize over $M\times M$ bipartite maximally entangled states $\ket{\Ps_M}$.
\el
An equivalent statement is presented in \cite[Proposition 2.4.12]{functional}.
That is if $\bra{\Ps_M} \r \ket{\Ps_M} > {k \over N}$ for some maximally entangled state $\ket{\Ps_M} $ then $\sn(\r)>k$. This result can be used to infer the Schmidt number of quantum states. For example let us consider the mixed state $\r=\sum_i p_i \proj{\ps_i}$, where $\ket{\ps_1}$ has the maximum Schmidt rank. 
The greater $p_1$ is, the greater $\bra{\Ps_M} \r \ket{\Ps_M}$ becomes. Then Lemma \ref{le:th00} shows that the Schmidt number of $\r$ also increases.

\subsection{Positive map}
\label{subsec:pm}

In this subsection, we investigate the Schmidt number in the view of positive and copositive maps. They play the fundamental roles in operator algebra and have a deep connection with quantum information. For example, the known Peres-Horodecki criterion says that a two-qubit or qubit-qutrit state is separable if and only if its partial transpose is a positive-semidefinite matrix. Here the transpose is a positive but not $2$-positive map. In general we define the positive and copositive maps as follows.
\bd
\label{df:kpositive}
A map $\phi\in B(M_m(\mathbb{C}),M_n(\mathbb{C}))$ is said to be $k$-positive/$k$-copositive if the map $id_k\ox \phi/\tau_k\ox \phi$ is positive, respectively.
\ed
Here $\tau_k$ is the transpose map in $B(M_k(\mathbb{C}),M_k(\mathbb{C}))$. Denote by $P_k[m,n]/P^k[m,n]$ the set of all $k$-positive/$k$-copositive maps in $B(M_m(\mathbb{C}),M_n(\mathbb{C}))$. Using these definitions we introduce completely positive and completely copositive maps.
\bd
\label{df:cp}
A map $\phi\in B(M_m(\mathbb{C}),M_n(\mathbb{C}))$ is completely positive/completely copositive if for every positive integer $k$, $\phi$ is $k$-positive/$k$-copositive, respectively. $\phi$ is said to be decomposable if it is the sum of a completely positive map and a completely copositive map.
\ed
With the well known dual cone relation \cite{stm11,itoh86,ek2000,facial} between positive maps and quantum states, the Schmidt number of an $m\times n$ entangled state $\r$ can be rephrased as 
\begin{subequations}
\label{eq:sn-pos}
\begin{align}
\sn(\r)&=\max_{l}\{l:\exists\ \phi\in P_l\ \text{s.t.}\ \tr(\r C_\phi^t)<0\}+1,\label{eq:sn-posa}\\
       &=\min_{l}\{l:\tr(\r C_\phi^t)\geq0\ \forall\ \phi\in P_l\}\label{eq:sn-posb}.	
\end{align}
\end{subequations}
Here $C_\phi=\sum_{i,j=1}^{m}\ketbra{i}{j}\ox\phi(\ketbra{i}{j})$ is the Choi matrix of the positive map $\phi$, and $C_\phi^t$ is the transpose of $C_\phi$. We denote the pairing of a quantum state $\r$ and a positive map $\phi$ by $\langle\r,\phi\rangle=\tr(\r C_{\phi}^t)$. If $\langle\r,\phi\rangle<0$, then $\phi$ is called an entanglement witness by which one can detect whether $\r$ is entangled \cite{terhal2000}.  If such a map exists, then the detected state has Schmidt number at least two. To decide the Schmidt number of $\r$, one should continue to test $\r$ using $k$-positive maps as entanglement witnesses until for certain $k$, no $k$-positive map can serve as an entanglement witness to $\r$. Let us illustrate this principle by assuming that $\r$ is a $3\times3$ entangled PPT state. We will make use of the following result from \cite{ylt16}:
\bl
\label{le:ylt}
Every 2-positive or 2-copositive map in $B(M_3(\mathbb{C}),M_3(\mathbb{C}))$ is decomposable. 
\el
The Lemma implies that 
for any $2$-positive map $\phi\in B(M_3(\mathbb{C}),M_3(\mathbb{C}))$, $\phi=\phi_1+\phi_2$ and $\phi_1/\phi_2$ is completely positive/completely copositive, respectively. Then $\tr(\r C_\phi^t)=\tr(\r C_{\phi_1}^t)+\tr(\r C_{\phi_2}^t)=\tr(\r C_{\phi_1}^t)+\tr(\r^\G (C_{\phi_2}^t)^\G)=\tr(\r C_{\phi_1}^t)+\tr(\r^\G (C_{\phi_2}^\G)^t)\geq 0$ because all matrices involved are positive. By \eqref{eq:sn-posb} every $3\times3$ PPT entangled state $\r$ has Schmidt number 2 since no $2$-positive map can serve as an entanglement witness to them.
As another application of \eqref{eq:sn-pos}, next we show that the Schmidt number is stable under perturbation. 
\bl
\label{le:SNstb}
For any bipartite states $\r$ and $\s$ with $\sn(\s)\leq\sn(\r)$, the Schmidt number of the perturbation $\r+\epsilon\s$ remains $\sn(\r)$ for sufficiently small $\epsilon>0$.
\el
\bpf
For any $l\geq \sn(\r)\geq \sn(\s)$, we have $tr(\r C_{\phi}^t)\geq0\ \forall \phi\in P_l$ and $tr(\s C_{\phi}^t)\geq0\ \forall \phi\in P_l$ by equation \eqref{eq:sn-posb}. Therefore $tr((\r+\epsilon\s) C_{\phi}^t)=tr(\r C_{\phi}^t)+\epsilon tr(\s C_{\phi}^t)\geq0\ \forall \phi\in P_l$ for any non-negative $\epsilon$. On the other hand, taking $l=\sn(\r)-1$, there exists a positive map $\psi\in P_l$ such that $tr(\r C_{\psi}^t)<0$ by equation \eqref{eq:sn-posa}. Choosing a sufficiently small $\epsilon$, we also have $tr((\r+\epsilon\s) C_{\psi}^t)=tr(\r C_{\psi}^t)+\epsilon tr(\s C_{\psi}^t)<0$. Hence by equation \eqref{eq:sn-posb} we have $\sn(\r+\epsilon\s)$ remains $\sn(\r)$ for sufficiently small $\epsilon$.
\epf

A similar property holds for quantum entanglement. That is, if $\r$ is entangled, then $\r+\e\s$ remains entangled for sufficiently small $\epsilon>0$. This fact can be proved by using the entanglement witness.



Let us recall the \textit{reduction map} $\L(\a)=(\tr\a) I - \a$ for any positive semidefinite matrix $\a$ \cite{hh1999}. Let $\L_A$ and $\L_B$ be the maps respectively acting on the system $A$ and $B$. One can show
\bea
\label{eq:reduction}
\L_A(\r)=I_A\ox \r_B-\r,
\notag\\
\L_B(\r)=\r_A\ox I_B-\r,
\eea
for any bipartite state $\r$. The reduction map is a positive but not completely positive (PNCP) map. If both matrices in \eqref{eq:reduction} are semidefinite positive then we say that $\r$ satisfies the reduction criterion. Otherwise $\r$ violates the reduction criterion, i.e., one of the two matrices in \eqref{eq:reduction} is not semidefinite positive. It is known that if the reduction criterion is violated then $\r$ is distllable \cite{hh1999}. The reduction criterion is weaker than the PPT criterion.



\subsection{Linear algebra}
\label{subsec:la}

In this subsection we review and construct a few results on linear algebra used throughout the paper. We have seen in Definition \ref{df:sn} that computing the Schmidt number of a quantum state requires the investigation of all decompositions the state. The following result provides the closed formula for the decomposition \cite{hjw1993}.
\bl
\label{le:sd}
Let $\r$ be a quantum state and the spectral decomposition $\r=\sum_i p_i \proj{a_i}$ such that $p_i>0$ and the $\ket{a_i}$ are pairwise orthonormal states. Then any decomposition $\r=\sum^m_{j=1} q_j \proj{b_j}$ with $q_j>0$ satisfies $\sqrt{q_j}\ket{b_j}=\sum_i u_{ij} \sqrt{p_i} \ket{a_i}$ for an order-$m$ unitary matrix $[u_{ij}]$.
\el
The Lemma will be used in the proof of Lemma \ref{le:proj} studying the Schmidt number of quantum states and their projection.
The next result is used for detecting the Schmidt number of bipartite states in Lemma \ref{le:>2}.

\bl
\label{le:psi+phi}
Suppose $\ket{\ps}$ and $\ket{\ph}$ are two bipartite states in $\cH_A\ox\cH_B$. There exists a nonzero state $\ket{\g}\in\cH_A$ or $\cH_B$ such that the two states $\braket{\g}{\ps}$ and $\braket{\g}{\ph}$ in $\cH_B$ or $\cH_A$ are proportional, and one of them is nonzero.
\el
\bpf
Suppose $\{\ket{a_j}\}_{j=1,\cdots,M}$ and $\{\ket{b_j}\}_{j=1,\cdots,N}$ are respectively two orthonormal basis in $\cH_A$ and $\cH_B$. Without loss of generality, we assume that $\ket{\psi}$ is not parallel to $\ket{\phi}$. 
We write the Schmidt decomposition as $\ket{\ps}=\sum^L_{j=1} c_j \ket{a_j,b_j}$ where $c_j\ne0$, $L\le M\le N$, and $\ket{\ph}=\sum^M_{j=1} \sum^N_{k=1} d_{jk} \ket{a_j,b_k}$. If some $d_{jk}\ne0$ when $L<j$ or $L<k$, then we choose $\ket{\g}=\ket{a_j}$ or $\ket{\g}=\ket{b_k}$, and the assertion holds. If all $d_{jk}=0$ when $L<j$ or $L<k$, 
we can find two complex number $x,y$ such that the nonzero state $x\ket{\ps}+y\ket{\ph}$ has Schmidt number strictly less than $L$. Choose $\ket{\g}\in\lin\{\ket{a_1},\cdots,\ket{a_L}\}$ and $\bra{\g}(x\ket{\psi}+y\ket{\phi})=0$. Then the two states $\braket{\g}{\ps}$ and $\braket{\g}{\ph}$ in $\cH_B$ are proportional, and $\braket{\g}{\ps}$ is nonzero.
So the assertion holds. 
This completes the proof.
\epf

Note that the space in which the state $\ket{\g}$ belongs to cannot be fixed. An example is that $\ket{\psi}=\ket{00}+\ket{11}$ and $\ket{\phi}=\ket{01}+\ket{12}$. 
One can show that no $\ket{\g}\in\cH_A$ satisfies the assertion. On the other hand one can choose $\ket{\g}=\ket{0}\in\cH_B$.






\section{Schmidt number of bipartite states}
\label{sec:main}

In this section we investigate the Schmidt number of bipartite states under local projections. Bipartite entangled states are the fundamental resources in quantum computing and cryptography. For this purpose bipartite states are converted into Bell states with a smaller Schmidt number under local projections asymptotically. This is the well-known entanglement distillation or purification \cite{bds96}. Next, bipartite states are entangled if and only if they have Schmidt number greater than one. Deciding whether a state is entangled is the well-known separability problem. One may detect the entanglement by locally projecting the target state onto another state with smaller dimensions. The local projections play important roles in both issues. We begin by proposing a preliminary Lemma on the Schmidt number and local projections.

\bl
\label{le:proj}
Let $\r$ be an $M\times N$ entangled state, $k\in[1,M-1]$ an integer, $P$ a matrix of rank $M-k$, and $\s=(P\ox I_B)\r(P^\dg \ox I_B)$ the projected state. Then
\\
(i) 
\bea
\label{eq:snr-k}
\max\{1,\sn(\r)-k\} \le \sn(\s) \le \min\{\sn(\r),M-k\}.
\notag\\
\eea
\\
(ii) We have
 $\r=\sum_j \proj{\ps_j}$, where $\ket{\ps_j}=\sum^{\sn(\s)}_{l=1} \ket{a_{j,l},b_{j,l}} + \sum^{k}_{i=1} \ket{z_i,y_{j,i}}$, $\cR(P)=\lin\{\ket{a_{j,l}}\}$, $\ket{z_i}\perp\ket{z_j}$, and $\ket{z_i}\perp P$ for all $i,j$.
\\ 
(iii) If $\sn(\r)=M$, then $\sn(\s)=M-k$.
 
Below we further assume that $\r$ is PPT. Then
\\
(iv)  
\bea
\label{eq:snr-k-gamma}
\max\{1,\sn(\r^\G)-k\} \le \sn(\s^\G) \le \min\{\sn(\r^\G),M-k\}.
\notag\\
\eea
\\
(v) If $k=\sn(\s)=1$, then $\sn(\r)=\sn(\r^\G)=2$.
\\
(vi) If $k=\min\{\sn(\r),\sn(\r^\G)\}-s$, and $\sn(\r)\ne\sn(\r^\G)$, then $\max\{\sn(\s),\sn(\s^\G)\}\geq s+1$.
\\
(vii) If $k=M-2$ or $M-1$, then $\sn(\s)=\sn(\s^\G)\in\{1,2\}$. 
\\
(viii) If $\sn(\r)=\sn(\r^\G)$, then $\sn(\s)-\sn(\s^\G)\in[-k,k]$.
\el
\bpf
(i)
Since the Schmidt number of quantum states is invariant up to local invertible operators,
we may assume that $P$ is a projector. Let $P=\sum^{M-k}_{i=1} \ketbra{v_i}{v_i}$ and $\{\ket{v_1},\cdots,\ket{v_M}\}$ an o. n. basis of $\cH_A$. Let $\r=\sum_j \proj{\ps_j}$ where $\ket{\ps_j}=\sum^M_{i=1} \ket{v_i,u_{ij}}$ and $\ket{u_{ij}}$ are nonnormalized vectors. 
We have
\bea
\label{eq:poxib}
\ket{\a_j}
&:=&(P\ox I_B)\ket{\ps_j}
\notag\\
&=&\sum^{M-k}_{i=1} \ket{v_i,u_{ij}}
:=\ket{\ps_j}-\ket{\b_j},
\eea
where 
\bea
\label{eq:bj}
\ket{\b_j}=\sum^{M}_{i=M-k+1} \ket{v_i,u_{ij}}.
\eea
Using Lemma \ref{le:sd} we may assume that $\ket{\a_j}$ are pairwise orthogonal, and we do not change the expression of $\r$ since there is no confusion.
Since $\s=\sum_j \proj{\a_j}$, we can find a unitary matrix $W=[w_{jl}]$ such that  for any $k$
the pure state $\sum_j w_{jl} \ket{\a_j}$ has Schmidt rank at most $\sn(\s)$.
Hence
\bea
\r
&=&\sum_j (\ket{\a_j}+\ket{\b_j})(\bra{\a_j}+\bra{\b_j})
\notag\\
&=&\sum_l
\bigg( \sum_j w_{jl} (\ket{\a_j}+\ket{\b_j}) \bigg)
\bigg( \sum_j w_{jl} (\bra{\a_j}+\bra{\b_j}) \bigg).
\notag\\
\eea
The definition of Schmidt number and \eqref{eq:bj} imply that $\sn(\r)\le \sn(\s)+k$. Since $\s$ is nonzero we always have $\sn(\s)\ge1$. So we have proved the lower bound in \eqref{eq:snr-k}.

On the other hand,
it is known that the Schmidt number is non-increasing under the local operations and classical communications \cite{th00}. So $\sn(\s)\le\sn(\r)$. Besides, the inequality $\sn(\s)\le M-k$ follows from the fact that $P$ has rank $M-k$. 
We have proved (i).

(ii) It suffices to prove $\cR(P)=\lin\{\ket{a_{j,l}}\}$. The inclusion $\cR(P)\supseteq\lin\{\ket{a_{j,l}}\}$ is evident. If the inclusion is strict, then
$\rank P>\rank \s_A$.

On the other hand Since $(P\ox I_B)\r(P^\dg\ox I_B)=\s$, we have $P\r_A P^\dg =\s_A$. Since $\rank \r_A=M$ we have $\rank P = \rank\s_A$. We have a contradiction and thus $\cR(P)=\{\ket{a_{j,l}}\}$.

(iii) The assertions both follow from the proof of (i).

(iv) The assertion follows from (i) by replacing $\r$ by $\r^\G$. 

(v) Since $k=1$ and $\sn(\s)=1$, (i) implies $1\le \sn(\r) \le 2$, and (iv) implies $1\le \sn(\r^\G) \le 2$. Since $\r$ and $\r^\G$ are both separable or not, we have proved the assertion. 


(vi) The assertion follows from (i).

(vii) The assertion follows from (i).

(viii) The assertion follows by summing up \eqref{eq:snr-k} and minus $\eqref{eq:snr-k-gamma}$. This completes the proof. 
\epf

By checking the proof of Lemma \ref{le:proj}, one can show that it also holds when $M>N$.
In Lemma \ref{le:proj} (i), the Schmidt number of the $M\times N$ bipartite state $\r$ is dominated by the sum of the Schmidt number of the projected states $\s$ plus the dimension of the kernel of the projection. 
In Lemma \ref{le:proj} (ii) if $\sn(\r)\le k$ then $\ket{\ps_j}=\sum^{k}_{i=1} \ket{z_i,y_{i,j}}$. It is impossible unless $k=M$. So the last inequality in \eqref{eq:snr-k} may be strict. 
An example  is the 1-undistillable $3\times3$ Werner state $\r$ and $k=1$. Since any $\s$ is a $2\times3$ state and still 1-undistillable, it is separable. So we have $\sn(\s)=1<\sn(\r)=2=M-k$. The first inequality in \eqref{eq:snr-k} may be also strict. First we give an example of NPT $\r$ and $M=N=3$. An example is the antisymmetric state $\r=\sum^2_{j,k=0,j<k} (\ket{jk}-\ket{kj})(\bra{jk}-\bra{kj})$. Up to ILOs we may assume the projector $P=\proj{0}+\proj{1}+(a\ket{0}+b\ket{1})\bra{2}$ where $a,b$ are complex numbers. Then $(P\ox I_2) \r (P^\dg \ox I_2)$ is an  NPT two-qubit state for any $a,b$. So it is entangled, and $\sn(\r)=\sn(\s)=2$.
Below is an example of PPT state, where $k=1$ and $\sn(\r)=2$. Note that these two states also saturate the last equality in \eqref{eq:snr-k}.

\bex
\label{ex:srho=ssigma=2}
Let $\r=\a\op\b$ be a PPT entangled state, where $\a$ and $\b$ are both $3\times3$ PPT entangled states, $\cR(\a_A)=\cR(\a_B)=\lin\{\ket{1},\ket{2},\ket{3}\}$ and $\cR(\b_A)=\cR(\b_B)=\lin\{\ket{4},\ket{5},\ket{6}\}$. It follows from Lemma \ref{le:dsum} and Corollary \ref{cr:3x3} that $\sn(\r)=\sn(\a)=\sn(\b)=2$. 
  
Let $P$ be a projector of rank five on $\cH_A$. We can express $P$ as
$
P=\sum^6_{i=1} \ketbra{a_i}{i},
$
where $\ket{a_1},\cdots,\ket{a_6}$ span a 5-dimensional subspace in $\bbC^6$. Hence either $\ket{a_1},\ket{a_2},\ket{a_3}$ or $\ket{a_4},\ket{a_5},\ket{a_6}$ span a 3-dimensional subspace in $\bbC^6$.
Let $\s=(P_A\ox I_B)\r(P_A\ox I_B)$. We have
\bea
\label{eq:sigma}
\s&=& 
\bigg(\sum^3_{i=1} \ketbra{a_i}{i}\bigg)_A
\a
\bigg(\sum^3_{i=1} \ketbra{i}{a_i}\bigg)_A
\notag\\
&\op_B&
\bigg(\sum^6_{i=4} \ketbra{a_i}{i}\bigg)_A
\b
\bigg(\sum^6_{i=4} \ketbra{i}{a_i}\bigg)_A.
\eea
So either the first state or the second state in  \eqref{eq:sigma} is still a $3\times3$ PPT entangled state. It follows from Lemma \ref{le:dsum} and Corollary \ref{cr:3x3} that $\sn(\s)=2=\sn(\r)$. 
\qed
\eex

In Lemma \ref{le:proj} (iii), one can generate quantum states of Schmidt number $M-k$  using rank $M-k$ projections from a Schmidt number $M$ state. The converse of (iii) does not hold. An example is the normalized antisymmetric projector on the $3\times3$ subspace. This is an entangled state. Further we propose an example of separable state. Consider a $2\times3$ PPT state $\r$ with any rank $1$ projection, we have $\sn(\r)=1<M$ and $\sn(\s)=1=M-k$.

Interestingly, Lemma \ref{le:proj} provides an alternative proof for a Conjecture in \cite{sbl2001}, see the Corollary below. An alternative proof using positive maps can be found in \cite{ylt16}.
\bcr
\label{cr:3x3}
Let $\r$ be a $3\times3$ state. Then
\\
(i) every PPT entangled $\r$ is of Schmidt number 2;
\\
(ii) every Schmidt-number-3 $\r$ is an NPT state. Moreover, for any matrix $P,Q\in M_3(\mathbb{C})$ with $rank(P)=rank(Q)=2$, the projected states $(P\ox I_3)\r(P^\dg\ox I_3)$ or $(I_3\ox Q)\r(I_3\ox Q^\dg)$ are NPT states. So $\r$ is distillable.
\ecr
\bpf
(i) This assertion follows Lemma \ref{le:proj} (i), in which we set $M=N=3$ and $k=1$. Then we have $\sn(\r)\leq\sn(\s)+1$. Note that $\s$ is a $2\times3$ PPT state which is also separable \cite{hhh96}.

(ii) The first assertion follows easily from (i). WLOG, assume that the projected states $\s=(P\ox I_3)\r(P^\dg\ox I_3)$ is a PPT state. So $\s$ is a separable state, hence it violates the inequality $\sn(\s)\geq \sn(\r)-k=2$. The last assertion follows from the fact that any $2\times N$ NPT states are distillable. This completes the proof.
\epf

The projected states may not be NPT even if the original state is NPT. For example, for any rank-one $P$ the state $(P\ox I_B)\r(P^\dg\ox I_B)$ is a separable state. It is an open problem to find out when the projected state is NPT, and it relates to the well-known distillability problem.
Next we consider the relation between the Schmidt numbers of the two tensors of the two copies of a bipartite state and the two copies of its projected state.
\bl
\label{le:sr2copy}
If $\r$ and $\s$ are introduced in Lemma \ref{le:proj}, then
\bea
\label{eq:2tensors}
\sn(\s^{\ox2})\leq\min\{\sn(\r^{\ox2}),(M-k)^2\},
\\
\label{eq:2tensors2}
\sn(\r^{\ox2})\leq\sn(\s)^2+2k\sn(\s)+k^2.
\eea
\el
\bpf
First we prove \eqref{eq:2tensors}. Since $\s=(P\ox I_B)\r(P^\dg \ox I_B)$, we can project $\r^{\ox2}$ onto $\s^{\ox2}$. Hence $\sn(\s^{\ox2})\le\sn(\r^{\ox2})$. It follows from \eqref{eq:snr-k} that
$\sn(\s) \le M-k$. So $\s$ is the convex sum of pure states of Schmidt rank at most $M-k$. So $\s^{\ox2}$ is the convex sum of pure states of Schmidt rank at most $(M-k)^2$. We have $\sn(\s^{\ox2})\le(M-k)^2$. So \eqref{eq:2tensors} holds. Next \eqref{eq:2tensors2} follows from the fact $\sn(\r)\le\sn(\s)+k$, which is from Lemma \ref{le:proj} (i) and (ii). This completes the proof.
\epf

The Lemma shows that the Schmidt number of the tensor product of the two copies of the same state is bounded by that of the tensor product of its projected states. One may similarly extend the Lemma to the tensor product of many copies of the same states. 
We further investigate the Schmidt number of the tensor product of different mixed states. The following result shows that such Schmidt number may be greater than the Schmidt number of each of them.

\bl
\label{le:>2}
Let $\r=\a_{A_1B_1}\ox\b_{A_2B_2}$ be a bipartite state on the system $A_1A_2$ and $B_1B_2$. 
\\
(i) If neither of the range of the states $\a_{A_1B_1}$ and $\b_{A_2B_2}$ contains any product state, then $\sn(\r)>2$, and any decomposition of $\r$ consists of pure states of Schmidt rank at least three.
\\
(ii) In (i) if $\sn(\r)=3$, then $\r=\sum_i \proj{\ps_i}$ where
\bea 
\label{eq:aibi}
\ket{\ps_i}
=
\ket{a_i}_{A_1A_2}\ket{b_i}_{B_1B_2}
+
\ket{c_i}_{A_1A_2}\ket{d_i}_{B_1B_2}
+
\ket{e_i}_{A_1A_2}\ket{f_i}_{B_1B_2},
\eea
is a bipartite state of Schmidt number three. For any $i$, the spaces $\cR((\r_i)_{A_1A_2})$ and $\cR((\r_i)_{B_1B_2})$ both have no product state.
\\
(iii) If $\a_{A_1B_1}$ and $\b_{B_1B_2}$ are both two-qutrit PPT entangled states of rank four, then $\sn(\r)=4$.
\el
\bpf
Since the range of the state $\a_{A_1B_1}$ does not contain any product state, $\a_{A_1B_1}$ is entangled. So $\r$ is also entangled and has Schmidt number at least two. 
Since the range of $\a_{A_1B_1}$ does not contain any product state, the pure state in any decomposition of $\r$ is a bipartite entangled state. 

We disprove the assertion. 
Suppose there is a decomposition of $\r$ containing a Schmidt-rank-two bipartite pure entangled state, i.e.,
$\r=\sum_i \proj{\ps_i}$ where
\bea 
\ket{\ps_1}=
\ket{a_1}_{A_1A_2}\ket{b_1}_{B_1B_2}+\ket{c_1}_{A_1A_2}\ket{d_1}_{B_1B_2}.
\eea It follows from 
Lemma \ref{le:psi+phi} that there exists a nonzero state $\ket{\g}\in\cH_{A_1}$ (or $\cH_{A_2}$) such that the two states $\braket{\g}{a_1}$ and $\braket{\g}{c_1}$ in $\cH_{A_2}$ (or $\cH_{A_1}$) are proportional, and one of them is nonzero. Hence $\braket{\g}{\ps_1}$ is a product state of the system $A_2$ (or $A_1$) and $B_1B_2$. By tracing out system $A_1B_1$ (or $A_2B_2$), we obtain that the range of $\b_{A_2B_2}$ (or $\a_{A_1B_1}$) contains a product state. It is a contradiction with the assumptions. So we have $\sn(\r)>2$, and any decomposition of $\r$ consists of pure states of Schmidt rank at least three.

(ii) The first assertion follows from (i). Using \eqref{eq:aibi} we shall regard $\ket{a_i},\ket{c_i},\ket{e_i}$ as an arbitrary basis of $\cR((\r_i)_{A_1A_2})$, and $\ket{b_i},\ket{d_i},\ket{f_i}$ as an arbitrary basis of $\cR((\r_i)_{B_1B_2})$. To prove the second assertion, it suffices to show that for any $i$, the states $\ket{a_i},\ket{b_i},\ket{c_i},\ket{d_i},\ket{e_i},\ket{f_i}$ all have Schmidt number greater than one. We have three cases.

In the first case, we assume that $\ket{a_i}$, $\ket{c_i}$ and $\ket{e_i}$ are product states. 
Let $\ket{a_i}=\ket{w_1,w_2}$, $\ket{c_i}=\ket{x_1,x_2}$ and $\ket{e_i}=\ket{y_1,y_2}$. The second assertion is trivial when for $j=1$ or $2$, two of the states $\ket{w_j},\ket{x_j}$ and $\ket{y_j}$ are proportional, or all of the three states are linearly independent. The only unsolved case is that for $j=1$ and $2$, any two of $\ket{w_j},\ket{x_j}$ and $\ket{y_j}$ are linearly independent and all of the three states are linearly dependent. According to Lemma \ref{le:psi+phi}, there exists a nonzero state $\ket{\g}\in\cH_{B_1}$ or $\cH_{B_2}$ such that the two states $\braket{\g}{d}$ and $\braket{\g}{f}$ in $\cH_{B_2}$ or $\cH_{B_1}$ are proportional, and one of them is nonzero. Let $\ket{z}\perp\ket{w_1}$ or $\ket{w_2}$, and $\ket{z}$ is not orthogonal to $\ket{y_1},\ket{z_1}$ or $\ket{y_2},\ket{z_2}$. Then $\braket{z,\g}{\ps_i}$ is a product state. 
We trace out $\r_{A_1B_1}$ by using the state $\ket{z,\g}$ as a state in the trace. Then one can show the second assertion, since the range of the state $\a_{A_1B_1}$ and $\b_{A_2B_2}$ does not contain any product state.

Next we assume that $\ket{a_i}$ and $\ket{c_i}$ are product states, and $\ket{e_i}$ is an entangled state. If $\ket{e_i}+x\ket{a_i}+y\ket{c_i}$ is a product state for some complex numbers $x,y$ then we have proved the assertion in the first case. So $\ket{e_i}+x\ket{a_i}+y\ket{c_i}$ is an entangled state for any $x,y$. It implies that there is a state $\ket{z}\in\cH_{A_1}$ or $\cH_{A_2}$ such that $\braket{z}{e_i}\ne0$ and $\braket{z}{a_i}=\braket{z}{c_i}=0$. By tracing out one of $\a_{A_1B_1}$ and $\b_{A_2B_2}$, we can obtain that the range of the other state contains product states. It is a contradiction with the assumption.
So we have proved the second assertion.

Third we assume that $\ket{a_i}$ is a product state, and $\ket{c_i}$ and $\ket{e_i}$ are both entangled states. If $\ket{e_i}+x\ket{a_i}+y\ket{c_i}$ is a product state for some complex numbers $x,y$ then we have proved the assertion in the last two cases. So $\ket{e_i}+x\ket{a_i}+y\ket{c_i}$ is an entangled state for any $x,y$. One can similarly show that $\ket{c_i}+x\ket{a_i}+y\ket{e_i}$ is an entangled state for any $x,y$. Lemma \ref{le:psi+phi} implies that there is a state $\ket{\g}\in \cH_{B_1}$ or $\cH_{B_2}$ such that the two states $\braket{\g}{d_i}$ and $\braket{\g}{f_i}$ in $\cH_{B_2}$ or $\cH_{B_1}$ are proportional, and one of them is nonzero. We have $\braket{\g}{\ps_i}=\ket{a_i}\ox\braket{\g}{b_i}+\ket{g_i}\ox\ket{h_i}$, where $\ket{g_i}$ is the linear combination of $\ket{c_i}$ and $\ket{e_i}$. So $\ket{g_i}$ is an entangled state. We can find a state $\ket{h}\in\cH_{A_1}$ or $\cH_{A_2}$ such that $\braket{h}{a_i}=0$ and $\braket{h}{g_i}\ne0$. So $\cR(\a_{A_1B_1})$ or $\cR(\b_{A_2B_2})$ contains a product state $\braket{h}{g_i}\ox\ket{h_i}$. It is a contradiction with the assumption.
So we have proved the second assertion.

One can similarly prove the second assertion by exchanging the systems $A_1A_2$ and $B_1B_2$.

(iii) It is known that neither of the range of the states $\a_{A_1B_1}$ and $\b_{A_2B_2}$ contains any product state. Further we can choose that $\ket{a_i}$ and $\ket{c_i}$ have Schmidt rank two, because $\cR(\r_{A_1A_2})$ is a 3-dimensional subspace of $\bbC^3\ox\bbC^3$.
Next if there is a state $\ket{\a}\in\cH_{A_1}$ or $\cH_{A_2}$ orthogonal to $\ket{a_i},\ket{c_i}$ and $\ket{e_i}$ at the same time, then $\cR(\r_{A_1A_2})\su \ket{a}^\perp\ox\bbC^3$. So $\cR(\r_{A_1A_2})$ contains a product state and it is a contradiction with (ii). Hence there is no state orthogonal to $\ket{a_i},\ket{c_i}$ and $\ket{e_i}$ at the same time. It implies that if there is a state $\ket{\a}\in\cH_{A_1}$ or $\cH_{A_2}$ orthogonal to $\ket{a_i},\ket{c_i}$, then there is a product state in $\cR(\a_{A_2B_2})$ or $\cR(\b_{A_1B_1})$. It is a contradiction with (ii). So such $\ket{\a}$ does not exist. We shall use these facts below.

It follows from Lemma \ref{le:psi+phi} that there exists a nonzero state $\ket{\g}\in\cH_{B_1}$ or $\cH_{B_2}$ such that the two states $\braket{\g}{d_i}$ and $\braket{\g}{f_i}$ in $\cH_{B_2}$ or $\cH_{B_1}$ are proportional, and one of them is nonzero. We have $\braket{\g}{\ps_i}=\ket{a_i}\ox\braket{\g}{b_i}+\ket{g_i}\ox\ket{h_i}$, where $\ket{g_i}$ is the linear combination of $\ket{c_i}$ and $\ket{e_i}$. We can find a state $\ket{h}\in\cH_{A_1}$ or $\cH_{A_2}$ such that $\braket{h}{a_i}=0$ and $\braket{h}{g_i}\ne0$. So $\cR(\a_{A_1B_1})$ or $\cR(\b_{A_2B_2})$ contains a product state $\braket{h}{g_i}\ox\ket{h_i}$. It is a contradiction with the assumption.
So we have proved the second assertion.
This completes the proof. \epf

Next we generalize Lemma \ref{le:>2} (i) to the tensor product of many bipartite states.

\bpp
\label{pp:snrho>n}
Let $\r=\ox^n_{j=1} \a_{A_jB_j}$ be a bipartite state of systems $A_1\cdots A_n:B_1\cdots B_n$, where $\a_{A_jB_j}$ are bipartite states of the system $A_jB_j$, $j=1,\cdots,n$, respectively. Suppose neither of $\cR(\a_{A_jB_j})$ contains any product state.
Then $\sn(\r)>n$, and  any decomposition of $\r$ consists of pure states of Schmidt rank at least $n+1$.
\epp
\bpf
By the definition of Schmidt number, it suffices to prove the second assertion. That is any decomposition of $\r$ consists of pure states of Schmidt rank at least $n+1$. Suppose it is wrong.
Let $\r=\sum_i \proj{\ps_i}$ where
\bea 
\label{eq:aibi}
\ket{\ps_1}
=
\ket{a_{1}}_{A_1\cdots A_n}\ket{b_{1}}_{B_1\cdots B_n}
+
\cdots
+
\ket{a_{k}}_{A_1\cdots A_n}\ket{b_{k}}_{B_1\cdots B_n},
\eea
is a bipartite pure state of Schmidt rank $k\le n$. Lemma 
\ref{le:psi+phi} implies that
there exists a nonzero state $\ket{\g}\in\cH_{A_1}$ such that the two states $\braket{\g}{a_1}$ and $\braket{\g}{a_2}$ in $\cH_{A_2\cdots A_n}$ are proportional, and one of them is nonzero. Let $\g'\in\cH_{B_1}$ be a state such that $\ket{\ph}:=\braket{\g,\g'}{\ps_1}\ne0$. So $\ket{\ph}$ is a bipartite pure state of Schmidt rank $k-1\le n-1$. Next using Lemma 
\ref{le:psi+phi} again, we can find a state $\ket{\d,\d'}\in\cH_{A_2B_2}$ such that $\braket{\b,\b'}{\ph}\ne0$ and has Schmidt rank at most $n-2$. Continuing in the same vein we can finally find a product state $\ket{\a}\in\cH_{A_1\cdots A_{n-1}:B_1\cdots B_{n-1}}$ such that $\braket{\a}{\ps_1}\in\cH_{A_nB_n}$ is nonzero and has Schmidt rank at most one. So it is a product state in $\cR(\r_{A_nB_n})$. This is a contradiction with the assumption. So we have proved $\sn(\r)>n$. This completes the proof. \epf

The result implies that there exists a PPT entangled state of Schmidt number $n$, where $n$ can be greater than any given integer. The state has equal birank $(r,r)$ for some integer $r$. Moreover, we can obtain a PPT entangled state of an arbitrary Schmidt number by the upcoming Lemma \ref{le:sym} from the aforementioned state.

\subsection{Approximation by Schmidt number}

Different quantum states may play the same role in quantum-information tasks. Their similarity decides how they play in the tasks.
The similarity of quantum states can be characterized by many quantum-information quantities, such as the fidelity, entanglement measure and equivalence under LOCC. 
In this subsection, we investigate the similarity between two quantum states in terms of their Schmidt number. First of all we present the following definitions.

\bd
\label{df:quantities}
Let $\r$ be an $M\times N$ entangled state, and $k\in[1,M-1]$ an integer. We define two quantities:
\bea
&&
\snmax(\r,k):=\max_P\{\sn(\s),
\notag\\&&
\s=(P\ox I_B)\r(P^\dg \ox I_B),
\dim\ker(P)=k\};
\\&&
\label{eq:min}
\snmin(\r,k):=\min_P\{\sn(\s),
\notag\\&&
\s=(P\ox I_B)\r(P^\dg \ox I_B),
\dim \ker(P)=k\}.
\eea
\qed
\ed
The two quantities in Definition \ref{df:quantities} can be estimated in a few special cases.
If $k=M-1$ then $\s$ is separable. We have $\sn_{\max}(\r,M-1)=\sn_{\min}(\r,M-1)=1$. If $k=M-2$ then we have $\sn_{\max}(\r,M-2),\sn_{\min}(\r,M-2)\in[1,2]$. One may similarly prove that  $\sn_{\max}(\r,1),\sn_{\min}(\r,1)\in[\sn(\r)-1,\sn(\r)]$.
Lemma \ref{le:proj} (i) implies that
\bea
\label{eq:snmax=snr}
\max\{1,\sn(\r)-k\}\leq \snmin(\r,k) \leq \snmax(\r,k) 
\notag\\
\leq \min\{\sn(\r),M-k\}.
\eea 
The condition by which $1 = \snmin(\r,k)$ or $\sn(\r)-k = \snmin(\r,k)$ holds is in Lemma \ref{le:proj} (ii). If 
$
\sn_{\max}(\r,k) = \sn(\r)
$ for some $k$,
then the space consisting all projected $\s$ best approximates $\r$ in terms of Schmidt number. It is difficult in general to determine whether such a best approximation exists for an abitrary $\r$.
The equalities depend on the dimensions $(M,N)$ as well as the pair $(\sn(\r),k)$. To illustrate, let $k=1$ and pick $\r$ from the set of all $3\times3$ PPT states. By Corollary \ref{cr:3x3} we know that $\sn(\r)=2$. Hence $1=\sn_{\max}(\r,1)<\sn(\r)=2$ since every $2\times3$ PPT states are separable. Consider $\r$ from the set of all $3\times3$ NPT states, then either $\sn(\r)=2$ or $\sn(\r)=3$. If $\sn(\r)=3$, by Corollary \ref{cr:3x3}, the projected states are NPT entangled states. Thus we have $2=\snmax(\r,1)<\sn(\r)=3$. If $\sn(\r)=2$, consider the antisymmetric state $\r=\sum^2_{j,k=0,j<k} (\ket{jk}-\ket{kj})(\bra{jk}-\bra{kj})$. Choose a projector $P=\proj{0}+\proj{1}$. Then $(P\ox I_2) \r (P^\dg \ox I_2)$ is entangled. The next Lemma shows the relation between the Schmidt number of a quantum state and its projection in terms of Definition \ref{df:quantities}.





\bl
\label{le:sn1=sn}
$\sn_{\max}(\r,k) = \sn(\r)$ holds for some $k$ if and only if $\sn_{\max}(\r,1) = \sn(\r)$.
\el
\bpf
The ``if'' part is trivial. It suffices to prove the ``only if'' part. Suppose $\snmax(\r,k) = \sn(\r)$. Since the Schmidt number does not increase under LOCC, we have
\bea
\snmax(\r,k) \le \cdots \le \snmax(\r,1) \le \sn(\r).
\eea
So the assertion holds. This completes the proof.
\epf

Note that $\snmax(\r,k)$ may not equal $\max_Q\{\sn(\s),\s=(I_A\ox Q)\r(I_A \ox Q^\dg), \dim \ker(Q)=k\}$. An example is $\r=\proj{\ps}+\proj{03}$, and $\ket{\ps}=\ket{00}+\ket{11}+\ket{22}$, $k=1$, $M=3$ and $N=4$. One can show that $\sn_{\max}(\r,1)=2$ and $\max_Q\{\sn(\s),\s=(I_A\ox Q)\r(I_A \ox Q^\dg), \dim \ker(Q)=1\}=3$. In general, we have the following Lemma.

\bl 
\label{le:sym}
Let $\r$ be an $M\times N$ entangled state, $P$ and $Q$ two nonzero projectors 
respectively on $\cH_A$ and $\cH_B$. Then
\\
(i) the following three integer sets are the same,
\bea
&&
\{\sn(\s):\s=(P\ox I)\r (P^\dg \ox I),~\forall P\ne0\}
\notag\\&=&
\{\sn(\s):\s=(I\ox Q)\r (I \ox Q^\dg),~\forall Q\ne0\}
\notag\\&=&
\{1,2,...,\sn(\r)\}.
\eea 
\\
(ii) For any $P$ there exists a $Q$ such that 
\bea
\sn\bigg( (P\ox I)\r (P^\dg \ox I) \bigg) 
=
\sn\bigg( (I\ox Q)\r (I \ox Q^\dg) \bigg).
\eea
\el
\bpf
(i) Consider the set $A_k=\{\sn(\s):\s=(P\ox I)\r (P^\dg \ox I),\dim \ker P\leq k\}$. By Lemma \ref{le:proj} (i), we obtain $A_1=\{\sn(\r)-1,\sn(\r)\}$ or $A_1=\{\sn(\r)\}$. Denote by $P_k$ a projector with $\dim \ker P_k=k$. Since any projection $P_k$ can be written into $P_k=P_1P_{k-1}$, we have $A_k=\{\sn(\s_{k}):\s_{k}=(P_1\ox I)\s_{k-1}(P_1^\dg \ox I),\s_{k-1}\in A_{k-1}\}$. Hence the set difference $A_k\backslash A_{k-1}$ is either an empty set or a set of single number by Lemma \ref{le:proj} (i). Using induction one has $A_{M-1}=\{1,...,\sn(\r)\}$. Similarly, we have the set $B_k=\{\sn(\s):\s=(I\ox Q)\r (I \ox Q^\dg),\dim \ker Q\leq k\}$ and $B_{N-1}=\{1,...,\sn(\r)\}=A_{M-1}$. 
\\
(ii) is an immediate consequence of (i).
\epf

We also conjecture that for $k=1,..,M-1$, the integer set $\{\sn(\s):\s=(P\ox I_B)\r(P^\dg \ox I_B), \dim \ker(P)=k\}$ is exactly the set of consecutive integers $\{\snmin(\r,k),...,\snmax(\r,k)\}$. The conjecure
holds when $k=M-1,M-2$ and $1$, as shown by the argument below \eqref{eq:min}. 
From Proposition \ref{pp:snrho>n} and Lemma \ref{le:sym}, we obtain a main result of this paper.
\bt
\label{thm:sr}
For any integer $r$, there exists a bipartite PPT entangled state of Schmidt number $r$.
\et



%

\section{Schmidt number of multipartite states}
\label{sec:multi}

Multipartite quantum states have a more complicated structure than that of bipartite states and have been extensively investigated in past years. For example the well-known $n$-partite Greenberger-Horne-Zeilinger (GHZ) state ${1\over\sqrt2}(\ket{0}^{\ox n}+\ket{0}^{\ox n})$ is the generalization of Bell state. It has been realized in experiments for small $n$ with a high fidelity and play an important role in quantum computing.
In this section we generalize the notion of Schmidt number to multipartite states.
The \textit{tensor rank}  of an $N$-partite quuantum state $\ket{\ps}\in\cH_1\ox\cdots\ox\cH_n$ of systems $A_1,\cdots,A_n$ is defined as the minimum integer $r$ such that there exist $r$ product states $\ket{a_{j,1},\cdots,a_{j,N}}$ and $\ket{\ps}=\sum^r_{j=1} \ket{a_{j,1},\cdots,a_{j,N}} $. For example the $n$-partite GHZ state has tensor rank two. Now Definition \ref{df:sn} can be generalized to multiipartite states as follows.
\bd
\label{df:tensorrank}
A multipartite density matrix $\r$􏰓 has Schmidt number $k$ if (i) for any decomposition of $\r$􏰓, $\{p_i>0, \ket{\ps_i}\}$ at least one of the vectors $\ket{\ps_i}$􏰃 has at least tensor rank $k$ and (􏰕ii)􏰀 there exists a decomposition of􏰓 $\r$ with all vectors $􏰂􏰥\ket{\ps_i}$􏰄􏰃 of tensor rank at most $k$.
\ed
For example, the three-qubit mixed state $\r=\proj{\a}+\proj{000}$ where $\ket{\a}=\ket{000}+\ket{111}$ has Schmidt number two. To understand this fact, we assume that $\r=\sum_i p_i \proj{\ps_i}$ as an arbitrary decomposition of $\r$. 
Using Lemma \ref{le:sd}, one can obtain that there is always some $\ket{\ps_i}$ of tensor rank two. Then Definition \ref{df:tensorrank} shows that $\sn(\r)=2$, and that the Schmidt number of multipartite states does not increase under LOCC. So the Schmidt number is also an entanglement measure for multipartite states.
Evidently, Definition \ref{df:tensorrank} reduces to Definition \ref{df:sn} for bipartite states $\r$. For simplicity we will regard tensor rank and Schmidt number as the same notion and use only Schmidt number. Further, the Schmidt number for bipartite and multipartite states are both invariant under ILOs. It is known that the Schmidt number is non-increasing under the local operations and classical communications \cite{th00}. So the Schmidt number is an entanglement monotone. Hence, the exact transformation under LOCC from a bipartite state $\ket{\ps}$ of smaller Schmidt rank to $\ket{\ph}$ of bigger Schmidt rank is impossible.
On the other hand, the transformation may be asymptotically realized by distilling EPR pairs from $\ket{\ps}$ and then preparing $\ket{\ph}$.
Third, it is known that for bipartite pure states $\ket{\ph}$ we have $\sn(\ket{\ph}^{\ox n})=n\sn(\ket{\ph})$. For multipartite pure states $\ket{\ps}$, we have $\sn(\ket{\ps}^{\ox n})\le n\sn(\ket{\ps})$ and the inequality is strict for the multiqubit W state $\ket{\ps}$ and integers $n>1$ \cite{ccm11}. 

In the following subsections we construct and investigate three quantities of multipartite states, namely the expansion, coarse graining and joint Schmidt number. Their definitions are respectively given in Definition \ref{df:expansion}, \ref{df:combine} and \ref{df:JSN}. The expansion describes the global states whose reduced density operators are the target multipartite states. The coarse graining constructs multipartite states from the known ones by combining systems. The joint Schmidt number is another Schmidt number of multipartite states  and different from Definition \ref{df:tensorrank}.
The main results are given in Theorem \ref{le:expansion}, Lemma \ref{le:combine}, Theorem \ref{le:jsn} and Lemma \ref{le:rank4}. These establish the connection between the Schmidt number, local ranks of reduced density operators and global multipartite states. 

\subsection{Expansion}
\label{subsec:exp}

In this subsection we investigate the Schmidt number of multipartite states and their reduced density operators. We review the notion of expansion which works for the well-known quantum marginal problem.

\bd
\label{df:expansion}
If $\r_A$ and $\r_B$ are the reduced density operators of a quantum state $\r_{AB}$, then we say that $\r_{AB}$ is an expansion of $\r_A$ and $\r_B$.
\ed
The expansion of a quantum state describes the global physical environment when the quantum state is regarded as a local state. When $\r_{AB}$ is a pure state, it is also called the purification of $\r_A$ and $\r_B$ in literatures.
For example if $\r_A=\r_B={1\over2}I_2$ then any two-qubit maximally entangled state $\r_{AB}$ is the expansion of $\r_A$ and $\r_B$. Some $\r_A$ and $\r_B$ do not have any purification (or even expansion). Using the definition we have
\bt
\label{le:expansion}
(i) The Schmidt number of $\r_{ABC}$ is not smaller than the Schmidt number of $\r_{AB}$, $\r_{AC}$ and $\r_{BC}$.
\\
(ii) $\r_{AB}$ has Schmidt number at most $k$ if and only if there is a tripartite state $\r_{ABC}$ of Schmidt number at most $k$.
\\
(iii) Suppose $\ket{\ps}_{ABC}$ is the purification of $\r_{AB}$. Then 
\bea
\label{eq:rank>=abc/ab>=1}
&&
\min\{\sn(\r_{AB}) \cdot \rank \r_{AB},~\rank\r_A\cdot\rank\r_B\}
\notag\\
\ge && \sn(\ket{\ps}_{ABC})
\notag\\
\ge && \max\{\rank\r_{AB},~\rank\r_A,~\rank\r_B\}
\notag\\
\ge &&
\sn(\r_{AB}).
\eea
\\
(iv) If $\r_{AB}$ is a PPT state, then the first two equalities in \eqref{eq:rank>=abc/ab>=1} hold simultaneously if and only if $\rank\r_A\cdot\rank\r_B=\rank\r_{AB}$ or $\sn(\r_{AB})=1$, i.e. $\r_{AB}$ is a separable state.
\\
(v) If $\r_{AB}$ is a PPT state then the three equalities in \eqref{eq:rank>=abc/ab>=1} hold simultaneously if and only if $\rank\r_A=\rank\r_B=1$.
\\
(vi) If $\r_{AB}=\proj{\ps}_{A_1B_1}\ox\sum_i \proj{ii}_{A_2B_2}$ is a bipartite NPT state where $\ket{\ps}=\sum_j \ket{jj}$, $A=A_1A_2$, $B=B_1B_2$, then the last equality in \eqref{eq:rank>=abc/ab>=1} holds. If $\r_{AB}$ has rank one then all three equalities in \eqref{eq:rank>=abc/ab>=1} hold.

\et
\bpf
(i) Let $\r_{ABC}=\sum_i\proj{\ps_i}$ where each $\ket{\ps_i}$ has Schmidt number at most $k:=\sn(\r_{ABC})$. So the pure states $\braket{i}{\ps_j}$ has Schmidt number at most $k$. Since $\r_{AB}=\tr_C \r_{ABC}=\sum_j\bra{j}_C\proj{\ps_i}\ket{j}_C$, the assertion on $\r_{AB}$ holds. The other assertions can be proved similarly.

(ii) The ``if'' part follows from (i). To prove the ``only if'' part, suppose $\r_{AB}=\sum_j\proj{\ps_j}_{AB}$ where each $\ket{\ps_j}$ has Schmidt number at most $k$. Then $\r_{ABC}=\sum_j\proj{\ps_j}_{AB}\ox\proj{j}_C$ is an expansion of $\r_{AB}$ and has Schmidt number at most $k$.

(iii) Suppose $\r_{AB}=\sum^l_{j=1} \proj{\a_j}_{AB}$ satisfies that $\sn(\a_j)\le \sn(\r_{AB})$. Without loss of generality, we may assume that the first $r:=\rank \r_{AB}$ states $\ket{\a_1},\cdots,\ket{\a_r}$ are linearly independent, and any $\ket{\a_j}$ is in the span of them. It is known  that $\ket{\ps}_{ABC}=\sum^l_{j=1}\ket{\a_j,u_j}$ where the $\ket{u_j}$'s form a set of o. n. basis in $\bbC^l$ \cite[Eq. (9.66)]{nc2000book}. Hence 
\bea
\sn(\ket{\ps}_{ABC})\le \sum^r_{j=1} \sn(\a_j) \le r \cdot \sn(\r_{AB}).
\eea
Next the inequality $\rank\r_A\rank\r_B\ge k:=\sn(\ket{\ps}_{ABC})$ follows from the definition of tensor rank. So we have proved the first inequality in \eqref{eq:rank>=abc/ab>=1}. Let $\r_{AB}=\sum^r_{i=1} \proj{\a_i}$ such that the $\ket{\a_i}$ are linearly independent. Then $\ket{\ps}_{ABC}=\sum^r_{i=1} \ket{\a_i,i}$, and thus $k\ge r$. Next the assertion $\sn(\ket{\ps}_{ABC})\ge\max\{\rank\r_A,\rank\r_B\}$ follows by writing $\ket{\ps}_{ABC}$ as the bipartite state of systems $A:BC$ and $B:AC$. So we have proved the second inequality in \eqref{eq:rank>=abc/ab>=1}. To prove 
the third inequality $\rank\r_A\ge \sn(\r_{AB})$ in \eqref{eq:rank>=abc/ab>=1}, we notice that $\r_{AB}=\sum_i p_i \proj{\ps_i}$ where each bipartite pure state $\ket{\ps_i}$ is an $M\times N$ state where $M\le\rank\r_A$ and $N\le\rank\r_B$.
So the inequality holds. 

(iv)
The "if" part can be verified straightforwardly. Next we prove the ``only if'' part. Since $\r_{AB}$ is a PPT state, then $\rank\r_{AB}\geq\max\{\rank\r_A,\ \rank\r_B\}$ \cite{hstt2003}. Hence the assumption of the ``only if'' part is equivalent to
\bea
&&
\min\{\sn(\r_{AB}) \cdot \rank \r_{AB},~\rank\r_A\cdot\rank\r_B\}
\notag\\
= && \sn(\ket{\ps}_{ABC})
\notag\\
= && \rank\r_{AB}.
\eea
If $\min\{\sn(\r_{AB})\cdot \rank \r_{AB},~\rank\r_A\cdot\rank\r_B\}=\sn(\r_{AB})\cdot \rank \r_{AB}$ then one obtains $(\sn(\r_{AB})-1)\cdot\rank(\r_{AB})=0$. Hence $\r_{AB}$ is separable. On the other hand if $\min\{\sn(\r_{AB})\cdot \rank \r_{AB},~\rank\r_A\cdot\rank\r_B\}=\rank\r_{A}\cdot\rank\r_{B}$ then it is obvious that $\rank\r_{A}\cdot\rank\r_{B}=\rank\r_{AB}$. 

(v) The assertion follows from (iv), \eqref{eq:rank>=abc/ab>=1} and $\rank\r_{AB}\geq\max\{\rank\r_A,\ \rank\r_B\}$.

(vi) The assertion can be verified straightforwardly using Lemma \ref{le:dsum}, because the states $\ket{\ps}_{A_1B_1}\ox \ket{jj}_{A_2B_2}$ are orthogonal each other.
This completes the proof.
\epf

When $k=2$, assertion (ii) gives a necessary and sufficient condition for whether $\r$ has Schmidt number at most two. Besides
the equality $\sn(\r_{ABC})=\sn(\r_{AB})=\sn(\r_{BC})=\sn(\r_{AC})$ may hold for some $\r_{ABC}$. An example is the three-qubit state $\ket{000}+\ket{a,a,a}$ where $\ket{a}=\ket{0}+\ket{1}$.
It is possible that 
\bea
\label{eq:monogamy}
\sn(\r_{A_1\cdots A_n})
&>&
\sum_{1\le j_1<j_2 \le n} \sn(\r_{A_{j_1} A_{j_2}})
\notag\\
&+&
\sum_{1\le j_1<j_2<j_3 \le n} \sn(\r_{A_{j_1} A_{j_2} A_{j_3} })
\notag\\
&+&
\cdots
\notag\\
&+&
\sum_{1\le j_1<\cdots<j_{n-1} \le n} \sn(\r_{A_{j_1} \cdots A_{j_{n-1}} }).
\eea 
For example, the inequality holds when $\r$ is the $d$-level Greenberger-Horne-Zeilinger state $\sum^d_{j=1} \ket{jj\cdots j}$ when $d$ is sufficiently big. The reason is that any $k$-partite reduced density operator $\s$ of $\r$ is a separable state, i.e., $\s=\sum_j p_j\proj{a_{j,1},\cdots,a_{j,k}}$. Hence $\sn(\s)=1$. All together we have $\sum_{k=2}^{n-1}\binom{n}{k}=2^n-\binom{n}{1}-\binom{n}{n}-\binom{n}{0}=2^n-n-2$ number of terms. If each system has dimension $d_k>2^n-n-2$, then any $d$ level GHZ state with $d>2^n-n-2$ will satisfy the inequality.
Since the Schmidt number is a multipartite entanglement measure, \eqref{eq:monogamy} shows the monogamy relation for some states. 

In Theorem (iii), we have shown the relation between the Schmidt number, the rank and the purification of a bipartite state. The known inequality $\rank\r_A\cdot\rank\r_B\ge\rank\r_{AB}$ holds for any state $\r_{AB}$. Eq. \eqref{eq:rank>=abc/ab>=1} gives the inequality $\rank\r_A\cdot\rank\r_B\ge \sn(\ket{\ps}_{ABC})\ge\rank\r_{AB}$ which is stronger than the known inequality.
In assertion (iv), if the state $\r_{AB}$ is not PPT then it may still make the first two equalities in \eqref{eq:rank>=abc/ab>=1} hold. For example $\r_{AB}$ is the bipartite pure entangled state.
A more complicated example is the mixed entangled state $\r_{AB}=\proj{\a}+\proj{\b}$ where $\ket{\a}=\ket{11}+\ket{22}$ and $\ket{\b}=\ket{33}+\ket{44}$. One can verify that the first two equalities in \eqref{eq:rank>=abc/ab>=1} holds since $\rank\r_A=\rank\r_B=4$, $\sn(\r_{AB})=\rank\r_{AB}=2$ and $\sn(\ket{\ps}_{ABC})=4$. On the other hand, the second equality in \eqref{eq:rank>=abc/ab>=1} fails when $\ket{\a}=\ket{01}+\ket{10}$ and $\ket{\b}=\ket{00}$. One can show that $\sn(\ket{\ps}_{ABC})=3>\sn(\r_{AB})=\rank\r_{AB}=2$. It is an interesting question to investigate when the last equality in \eqref{eq:rank>=abc/ab>=1} holds. 

For any tripartite state $\ket{\ps}_{ABC}$, if we regard it as a bipartite state over the split of systems $A$ and $BC$, then we obtain $\rank\r_A=\rank\r_{BC}$. Similarly one obtains $\rank\r_B=\rank\r_{AC}$, and $\rank\r_C=\rank\r_{AB}$. So only three of the six parameters $\rank\r_A,\rank\r_B,\rank\r_C,\rank_{AB},\rank_{AC},\rank_{BC}$ are independent. In fact we have chosen the three parameters $\rank\r_A,\rank\r_B$ and $\rank\r_{AB}$ in \eqref{eq:rank>=abc/ab>=1}. The other two parameters $\sn(\r_{AB})$ and $\sn(\ket{\ps}_{ABC})$ are also independent from the three parameters. On the other hand the six parameters of a mixed tripartite state may be independent from each other, and the investigation is more complicated.
For readers' reference, the relation between the ranks of global and local systems for the entropy has been recently investigated \cite{chl14}.

\subsection{Coarse graining}
\label{subsec:coa}

In this subsection we investigate the Schmidt number of multipartite states in terms of its coarse graining. The latter is defined as follows. 

\bd
\label{df:combine}
(i) Let $\r$ be an $n$-partite quantum state of systems $A_1$, $\cdots$, $A_n$. If we partition the systems into $m$ disjoint parties $B_1$, $\cdots$, $B_m$ then we obtain a new $m$-partite quantum state $\s$. We denote $\s$ as a corase graining of $\r$.
\\
(ii) The multipartite PPT states are defined as the states any bipartition of whom is a PPT state. We denote $\r^{\G_j}$ as the partial transpose w. r. t. system $A_j$.
\ed
For example if $\ket{\ps}=\ket{000}+\ket{111}$, $B_1=A_1$, and $B_2=A_2A_3$, then $\ket{\ph}=\ket{\ps}=\ket{00}+\ket{13}$ where $\ket{0}_{B_2}=\ket{00}_{A_2A_3}$ and $\ket{3}_{B_2}=\ket{11}_{A_2A_3}$. The following claim is clear from the definition.
\bl
\label{le:combine}
(i) The Schmidt number of a multipartite pure state is not smaller than that of its coarse graining.
\\
(ii) The multipartite state $\r$ and its partial transpose $\r^{\G_j}$ are simultaneously separable or not.
\el

We explain the coarse graining from the point of view of quantum information. In a multipartite state $\ket{\ps}$, some of the $n$ systems can be combined so that they perform collective operation, and create more quantum correlation quantitatively and qualitatively in $\ket{\ps}$. So the coarse graining of $\ket{\ps}$ represent different entanglement structure from $\ket{\ps}$. The coarse graining has been used to investigate the geometric measure of entanglement \cite{zch2010}.

\subsection{Joint Schmidt number}
\label{subsec:jsn}

In this subsection we construct another version of Schmidt number of multipartite states. This is different from Definition \ref{df:tensorrank}, namely the joint Schmidt number (JSN). We begin by reviewing the version of pure multipartite states constructed in \cite{hk16}.
\bd
\label{df:biSR}
If the multipartite state $\ket{\phi}\in \mathcal{H}_1\ox\cdots\ox\mathcal{H}_n$ has Schmidt number $s_l$ under the bi-partition $\mathcal{H}_l\ox(\ox_{j\neq l}\mathcal{H}_j)$, then we say that $\ket{\phi}$ has joint Schmidt number $\jsn(\phi)=(s_1,...,s_n)$.
\ed
For example, the genuinely entangled multiqubit state has joint Schmidt number $(2,\cdots,2)$. Essentially, the definition arises in the different bi-partitions of the systems. To generalize it to mixed multipartite states $\r$, we denote $\jsn(\r)$ as the joint Schmidt number of $\r$. Given two n-partite states $\r$ and $\s$ with $\jsn(\r)=(s_1,...,s_n)$ and $\jsn(\s)=(t_1,...,t_n)$, we say that $\s$ dominates $\r$ and denote it by $\jsn(\r)\le\jsn(\s)$ if $s_i\leq t_i$ for $i=1,...,n$. So two tuples $(s_1,\cdots,s_n)$ and $(t_1,\cdots,t_n)$ are equal when they dominate each other. 

\bd
\label{df:JSN}
The multipartite state $\r$ in the system $\prod_{i=1}^n A_i$ has joint Schmidt number $(s_1,\dots,s_n)$ if it has Schmidt number $s_l$ under the system bipartition of $A_l:\prod_{i\neq l} A_i$. If in addition there exists a decomposition $\r=\sum_i\proj{\phi_i}$ with all $\jsn(\ket{\phi_i})\le(s_1,\dots,s_n)$, then we say the decomposition is a balanced decomposition.
\ed
For example, the three-qubit state $\proj{\ps}+\proj{000}$ has joint Schmidt number $(2,2,2)$ where $\ket{\ps}=\ket{001}+\ket{010}+\ket{100}$. The definition implies that the multipartite state is separable if and only if it has a balanced decomposition with joint Schmidt number $(1,\dots,1)$. Furthermore, for any local operators $V=\ox^n_{j=1}V_j$, one can show that $\jsn(V\r V^\dg)\le \jsn(\r)$. Hence the joint Schmidt number is a multipartite entanglement monotone and is physically meaningful. This is similar to the role of Schmidt number for bipartite states. We further investigate the mathematical relation of them.
\bt
\label{le:jsn}
(i) Let $\ket{\ps}$ be a multipartite state of $\jsn(\ket{\ps})=(s_1,...,s_n)$. Then $\max_{j=1,\dots,n}\{s_j\}\le\sn(\ps)\le \min_{j=1,\dots,n}\{{\P^n_{i=1}s_i\over s_j}\}$.
\\
(ii) If $\ket{\ps}$ is separable under $(n-1)$ many bi-partitions, then $\ket{\ps}$ is separable.
\et
\bpf
(i) The lower bound $\max_{j=1,\dots,n}\{s_j\}\le\sn(\ps)$ follows from the definition of Schmidt number. We will prove the assertion that $\sn(\ps)\le \prod_{i\neq n}s_i$ and one can similarly prove the assertion. By definition we have 
$n$ ways of bipartition, namely $\ket{\ps}=\sum^{s_l}_{i=1} \ket{a^l_i}_{A_l}\ox\ket{b^l_i}_{\prod_{j\neq l}A_j}$ where $\ket{a^l_i}$ are orthonormal states and the superscript $l\in\{1,...,n\}$. 
Hence $\ket{\ps}=\sum^{s_l}_{i=1} \proj{a^l_i}_{A_i}\ket{\ps}$. By using this equation for $l=1,\dots,n-1$ we have
\bea
\ket{\ps}&=&
\ox^{n-1}_{l=1}\sum^{s_l}_{i=1} \proj{a^l_i}_{A_i}\ket{\ps}
\notag\\
&=&
\sum^{s_1}_{i_1=1} \cdots \sum^{s_{n-1}}_{i_{n-1}=1} 
\ket{a^1_{i_1},\cdots,a^{n-1}_{i_{n-1}}}_{A_1\cdots A_{n-1}}
\ket{\ps_{i_1,\cdots,i_{n-1}}},
\notag\\
\eea
where $\ket{\ps_{i_1,\dots,i_{n-1}}}=\braket{a^1_{i_1},\cdots,a^{n-1}_{i_{n-1}}}{\ps}$ is a vector in $\cH_n$.
So the assertion follows. 

(ii) The assertion follows from (i) immediately.
This completes the proof.
\epf

The bound in Theorem \ref{le:jsn} (i) is tighter than that in \cite[Theorem 4.2]{hk16}, which says $\sn(\r)\le\prod_{i=1}^ns_i$. For example consider the tripartite state $\ket{\ps}=\ket{111}+\ket{122}+\ket{213}+\ket{224}$. One can verify that $\sn(\ps)=4$ and $\jsn(\ket{\ps})=(2,2,4)$.
So $\sn(\ps)=s_1s_2<s_1s_2s_3=16$. On the other hand, any 4-partite pure state $\ket{\ph}_{A_1A_2A_3A_4}$ can be regarded as a tripartite state, say $\ket{\a}_{A_1,A_2,A_3A_4}$ in terms of Definition \ref{df:combine}. If $\jsn(\ket{\ph})=(s_1,s_2,s_3,s_4)$ then $\jsn(\ket{\a})=(s_1,s_2,s_3')$. So Lemma \ref{le:combine} says that $\sn(\ps)\ge\sn(\a)$, and Theorem \ref{le:jsn} says that $s_1s_2\ge \sn(\a)$. Hence
\bea
\min\{\sn(\ps),s_1s_2\}\ge\sn(\a).
\eea

The condition of $(n-1)$ many bipartitions in  Theorem \ref{le:jsn} (ii) is necessary. Indeed a multipartite state $\ket{\ps}$ may be entangled if its $(n-2)$ many bipartitions are all separable. An example is the tripartite state $\ket{\ps}=\ket{000}+\ket{110}$. In spite of Theorem \ref{le:jsn} (ii), the biseparability via all bi-partitions does not imply the separability of multipartite mixed states. An example is the 3-qubit PPT entangled state $\rho=I-\sum^4_{j=1}\ket{a_i,b_i,c_i}\bra{a_i,b_i,c_i}$ where $\{\ket{a_i,b_i,c_i}\}$ is a 3-qubit UPB. One can show that $\jsn(\r)=(1,1,1)$, and $\r$ has Schmidt rank two. Since $\sn(\r)=2>1^3/1=1$, Theorem \ref{le:jsn} (i) cannot be generalized to mixed states.

In fact, any multipartite PPT state of rank at most three, or any non-three-qubit and non-two-qutrit PPT state of rank four is separable \cite{cd13}. 
Thus it has joint Schmidt number $(1,1,\cdots,1)$. On the other hand, $\rho$ does not have a balanced decomposition, because $\rho$ is entangled. One can verify that for any $j=1,2,3$, $\r^{\G_j}$ is still a PPT entangled state of rank four, and satisfies $\jsn(\r^{\G_j})=\jsn(\r)=(1,1,1)$ and $\sn(\r^{\G_j})=\sn(\r)=2$. For general entangled states we propose the following statement.
\bl
\label{le:rank4}
Let $\r$ be a multipartite entangled PPT state of rank four. Then
\\
(i) $\r$ and its partial transpose w. r. t. any systems, when regarded as bipartite states, all have Schmidt number two.
\\
(ii) If $\r$ is not a two-qutrit state then $\jsn(\r)=(1,\cdots,1)$.
\\
(iii) Any multipartite entangled PPT state with Schmidt number at least 3 when regarded as bipartite states, has rank at least 5.
\el
\bpf
(i) It is known that any entangled PPT state $\r$ of rank four is either a three-qubit or a two-qutrit state \cite{cd13}. The assertion holds when $\r$ is a two-qutrit state by Corollary \ref{cr:3x3}. On the other hand if $\r$ is a three-qubit state, then $\jsn(\r)=(1,1,1)$ \cite{cd13}. So $\r$ is the convex sum of product states over the bipartition of spaces $\cH_1:\cH_{2,3}$. So the assertion also holds.

(ii) The assertion can be proved by the argument similar to that of (i).

(iii) Immediate from (i).This completes the proof.
\epf

Lemma \ref{le:rank4} (iii) restricts the rank of desired states whose Schmidt number is different from that of its partial transpose. So far there is no example or proof for the existence of such states.



\section{Problems}
\label{sec:con}

In this section we introduce some open problems on the Schmidt number.
Let $\r$ be a bipartite state, $P$ a projector on $\cH_A$, and $P^\perp$ the orthogonal projector to $P$. Let $\a=(P\ox I)\r(P\ox I)$ and $\b=(P^\perp\ox I)\r(P^\perp\ox I)$. Then it is natural that $\sn(\r)\le \sn(\a)+\sn(\b)$. However it is generally wrong and we give a counterexample. Let $\r=\proj{\ps}+\proj{\ph}+\proj{\o}$ where $\ket{\ps}=\ket{11}+\ket{22}$, $\ket{\ph}=\ket{33}+\ket{44}+\ket{55}$, and $\ket{\o}=\ket{33}-\ket{44}+\ket{66}$. Let $P=\proj{1}+\proj{3}+\proj{4}$. One can verify that $\a$ and $\b$ are both separable states. We claim that $\sn(\r)=3$ and thus the inequality is wrong. To prove the claim, we note that the maximal Schmidt rank of any state in $\cR(\r)$ is three, then the claim follows from the definition of Schmidt number and Lemma \ref{le:sd}.

Lemma \ref{le:sn1=sn} shows that
if $\snmin(\r,k) = \sn(\r)
$ or $\snmax(\r,k) = \sn(\r)
$ for some $k$, then the minimum $k$ is one. On the other hand $\snmin(\r,k) =\snmax(\r,k) = 1$ when $k=M-1$.
However
\bcj
(i) What is the maximum $j$, such that $\snmax(\r,j) = \sn(\r)$?
\\
(ii) What is the minimum $k$, such that $\snmax(\r,k) = 1$?
\ecj



\bcj
\label{cj:bsn}
(i) There exists a PPT state $\r$ such that $\sn(\r)>\sn(\r^\G)$.
\\
(ii) Such $\r$ exists in $M\times N$ system where $3\leq M\leq N$ and $MN\geq12$. The simplest $\r$ is a $3\times4$ PPT state of BSN $(2,3)$.
\\
(iii) If the simplest $\r$ in (ii) exists then $\sn(\r^{\ox 2})$ has BSN $(4,9)$.
\\
(iv) If (i) holds then there exists $\r$ constructed from a UPB $\{\ket{a_j,b_j}\}$, i.e., $\r=I-\sum_j \proj{a_j,b_j}$.
\ecj
Since Schmidt number is an entanglement measure, the equality $\sn(\r)=\sn(\r^\G)$ would imply that $\r$ and $\r^\G$ have the same entanglement. However, to find an example for Conjecture \ref{cj:bsn} (ii), one has to find a $3\times4$ entangled PPT state with Schmidt number 3 \cite{ylt16}. No concrete example has been given in the literature yet. The existence of a $3\times4$ PPT state $\r$ with $\sn(\r)=3$ is equivalent to the existence of an indecomposable 2-positive map in $B(M_3(\mathbb{C}),M_4(\mathbb{C}))$. Note that if such a state exists, then it may provide a candidate for an example for Conjecture \ref{cj:bsn}. One need to further check $\sn(\r^\G)=2$ besides $\sn(\r)=3$.



%
%


\bcj
For any positive integer $L$, there is a PPT state $\r$ such that $|\sn(\r)-\sn(\r^\G)|\geq L$.
\ecj

\bcj
If $\sn(\r)\ge\sn(\s)$, then $\sn(\r^{\ox 2})\ge \sn(\s^{\ox 2})$.
\ecj
If the conjecture holds, then $\sn(\r^{\ox 2^n})\ge\sn(\s^{\ox 2^n}), \forall n\geq1$ provided $\sn(\r)\geq\sn(\s)$.

\section*{Acknowledgments}

LC was supported by the NSF of China (Grant No. 11501024), and the Fundamental
Research Funds for the Central Universities (Grant Nos. 30426401, 30458601 and 29816133). WST was partially supported by Singapore Ministry of Education Academic Research Fund Tier 1 Grant (No. R-146-000-193-112).

\bibliographystyle{unsrt}

\bibliography{sn2}

\end{document}